\documentclass[%
reprint,
superscriptaddress,
nofootinbib,
nobibnotes,
 amsmath,amssymb,
 aps,
prl,
]{revtex4-1}

\usepackage[x11names]{xcolor}
\usepackage{graphicx}
\usepackage{dcolumn}
\usepackage{bm}
\usepackage{float}
\usepackage{tikz-cd}

\begin{document}


\title{Deep learning the Hohenberg-Kohn maps of Density Functional Theory}

\author{Javier Robledo Moreno}
\email{jrm874@nyu.edu}
\affiliation{
Center for Computational Quantum Physics, Flatiron Institute, New York, NY 10010 USA
}
\affiliation{Center for Quantum Phenomena, Department of Physics, New York University, 726 Broadway, New York, New York 10003, USA
}

\author{Giuseppe Carleo}
\email{gcarleo@flatironinstitute.org}
\affiliation{
Center for Computational Quantum Physics, Flatiron Institute, New York, NY 10010 USA
}

\author{Antoine Georges}
\email{ageorges@flatironinstitute.org}
\affiliation{
Center for Computational Quantum Physics, Flatiron Institute, New York, NY 10010 USA
}
\affiliation{Coll{\`e}ge de France, 11 place Marcelin Berthelot, 75005 Paris, France}
\affiliation{CPHT, CNRS, {\'E}cole Polytechnique, IP Paris, F-91128 Palaiseau, France}
\affiliation{DQMP, Universit{\'e} de Gen{\`e}ve, 24 quai Ernest Ansermet, CH-1211 Gen{\`e}ve, Suisse}

\date{\today}

\begin{abstract}
A striking consequence of the Hohenberg-Kohn theorem of density functional theory is the existence of a bijection between the local density and the ground-state many-body wave function. Here we study the problem of constructing approximations to the Hohenberg-Kohn map using a statistical learning approach. Using supervised deep learning with synthetic data, we show that this map can be accurately constructed for a chain of one-dimensional interacting spinless fermions, in different phases of this model including the charge ordered Mott insulator and metallic phases and the critical point separating them. However, we also find that the learning is less effective across quantum phase transitions, suggesting an intrinsic difficulty in efficiently learning non-smooth functional relations. We further study the problem of directly reconstructing complex observables from simple local density measurements, proposing a scheme amenable to statistical learning from experimental data.  
\end{abstract}

\maketitle

\begin{figure*}[tb]
        \includegraphics[width=1\textwidth]{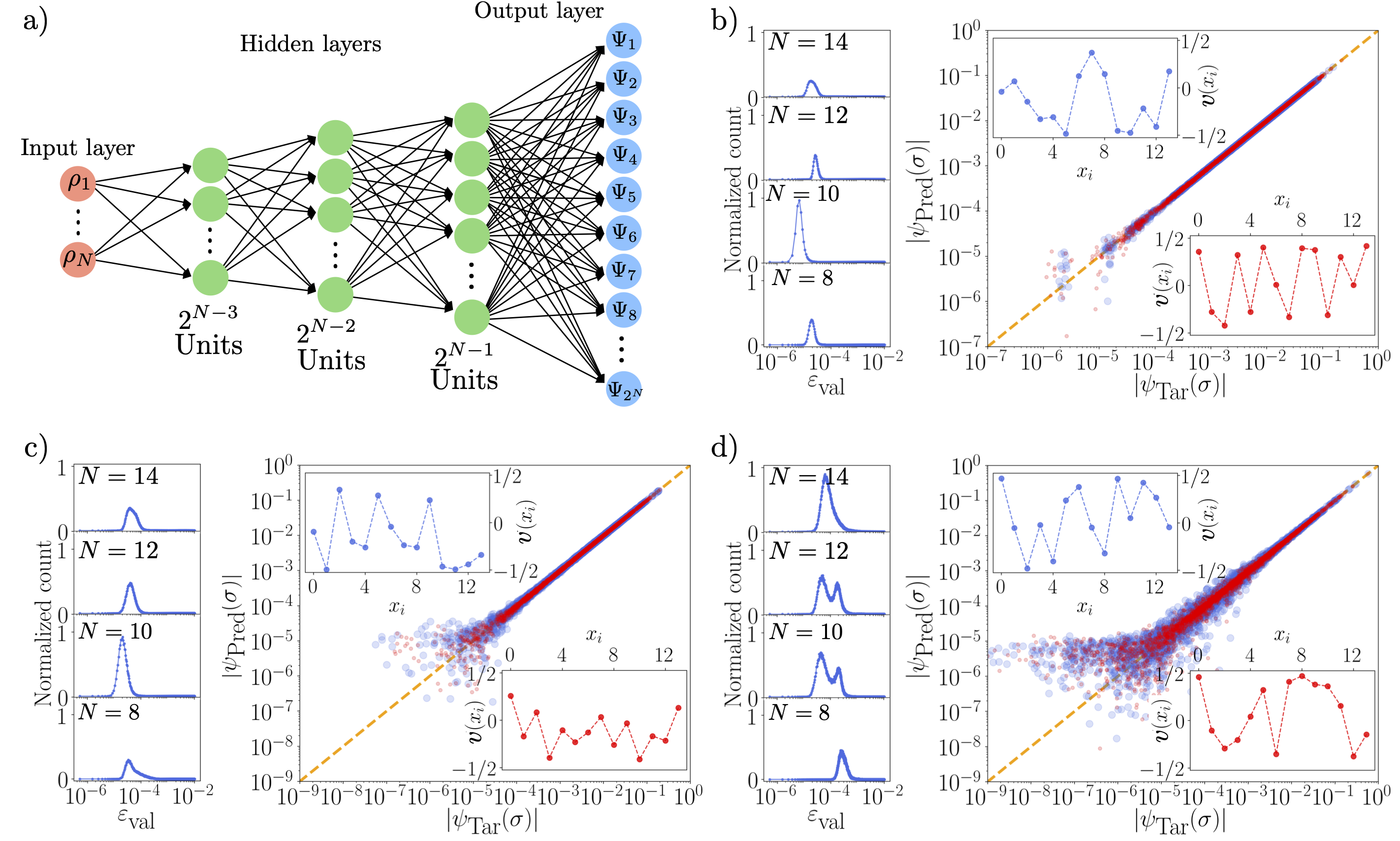}
        \caption{\label{FIG_01: Performance} \textbf{a)} Diagram of the neural network architecture used to represent the $\rho_i \rightarrow |\psi\rangle$ (DTWF) map. 
        \textbf{b)} $U/t = 1$ (metal),  \textbf{c)}  $U/t = 2$ (critical), \textbf{d)} $U/t = 4$ (Mott insulator). \textbf{Left:} Normalized histograms of the error functions as defined in Eq.~(\ref{Eq_06: Loss function}) over the validation set. \textbf{Right:} Predicted versus exact wave function components in the occupation number basis, given two different random potentials depicted in the insets. System size is $N = 14$. Red dots correspond to the potential in the inset in the lower right corner and blue ones to the potential in the upper left inset.}
    \end{figure*}

\paragraph*{Introduction.--}
    The Hohenberg-Kohn (HK) theorem~\cite{Hohenberg1964} is a founding principle of density functional theory (DFT). It establishes that there exists a bijective map between the local density and the one-body potential. This, remarkably, also implies a bijective relation between the local density and the ground-state many-body wave function of the system $\psi(\boldsymbol{r}_1,\boldsymbol{r}_2,\cdots,\boldsymbol{r}_N)$, which is thus a unique functional of the one-particle density $\rho(\boldsymbol{r})$. Hence, there is also an injective map connecting the local density with any observable of interest (such as two-point density correlation functions) in the ground state: 
     \begin{center}
     \begin{tikzcd}
        \rho(\boldsymbol{r}) \arrow{r}{\textrm{DTWF}} \arrow[black, bend right]{rr}[black,swap]{\textrm{DTCF}}  & \psi(\boldsymbol{r}_1, \boldsymbol{r}_2, ...) \rar \arrow{l}  & \langle n(\boldsymbol{r})n(\boldsymbol{r}')\rangle_{\psi}
    \end{tikzcd}
     \end{center}
     where DTWF is the density to ground-state wave function map and DTCF labels the density to two-point density correlator map. 
     However, the exact form of the DTWF and DTCF maps is unknown in most cases \cite{Cohen2012, KohnNobelLecture} and phenomenological 
     approximations are required to construct them. These approximations typically lead to inaccurate predictions when the electrons are strongly correlated~\cite{Cohen792, Mori2008}, as in Mott insulating phases.  
     
     With the recent interest in machine learning (ML) techniques applied to physical sciences~\cite{Carleo2019}, data driven approaches have successfully been applied to DFT for different applications. Some works use ML techniques in the Kohn-Sham (KS)  scheme \cite{Kohn1965}, to improve or parametrize exchange-correlation functionals and potentials~\cite{Kolb2017,Liu2017,Lundgaard2016, Nudejima2019, Schmidt2019, Nagai2018, Dick2019}, the non interacting kinetic energy functionals and their derivatives~\cite{Golub2019, Kolb2017, Seino2018, Snyder2013, Snyder2016} or the full density functional~\cite{Snyder2016, Snyder2012,Li2016}. Other works take a more direct approach based on the HK theorem to learn the potential to density map and the potential to ground-state energy map~\cite{Brockherde2017}, potential to energy spectrum map~\cite{Pilati2019}, infer relevant energies of the system from the local density~\cite{Nelson2019} or the external potential~\cite{Ryczko2019}. Despite this progress, little is known about the practical computational complexity of the statistical learning of the direct DTWF and DTCF HK maps, and in what regimes the learning approach can fail.
    
    In this article, we systematically investigate the problem of reconstructing the ground-state wave function (DTWF) and the correlation functions (DTCF) from the knowledge of the local density, using supervised deep learning. Focusing on a lattice model of interacting electrons, we show that the DTWF map can be learned for different phases of the model, including Mott and metallic phases and the critical point. The explicit ML representation of the DTWF map allows for the possibility to gain insight into this high dimensional bijection and its properties in models with finite basis sets such as small molecules. In particular, we find that learning the map through a  quantum phase transition (QPT) leads to intrinsic representational difficulties.
    Finally, we show that the DTCF map allows one to compute physical quantities of interest --like two-point correlators--  and their scaling laws directly from the local density of the system. Specifically, we could extract the Luttinger liquid (LL) parameter, finite-size scaling behaviour and logarithmic corrections from the ML-constructed correlation functions. This opens the possibility of reconstructing non-trivial physical quantities directly from experimental X-ray measurements of the electron density~\cite{Grabowsky2017, Tolborg2019}, or learning these maps from quantum simulation experiments~\cite{Lye2005, White2009, Celi2019, Omran2019, Keesling2019, Bloch2008, Scherg2018, Sherson2010}.

\paragraph*{The Hohenberg-Kohn Theorem.--}
    The fundamental theorem of DFT is formulated in the context of interacting electrons subjected to an external potential $v(\boldsymbol{r})$, whose Hamiltonian has the form:
    \begin{equation}\label{Eq_00: Electron Hamiltonian}
        \hat{H} = \hat{K} + \hat{U} + \hat{V},
    \end{equation}
    where $\hat{K}$ is the kinetic energy operator, $\hat{U}$ are the two-body interactions and $\hat{V}$ is the one-body external potential. For fixed  $\hat{K}$ and $\hat{U}$, 
    the HK theorem states that there is a one-to-one correspondence between the local electron density in the ground-state $\rho(\boldsymbol{r})$ and the external potential $v(\boldsymbol{r})$.
     
    Given the generality of this theorem, we consider in this work a Hamiltonian which is simpler than the full inhomogeneous electron gas, 
    but has a structure similar to Eq.~(\ref{Eq_00: Electron Hamiltonian}), namely a 1D extended Hubbard model of spinless Fermions in a lattice of $N$ sites with periodic boundary conditions:
    \begin{equation}\label{Eq_01: Full hamiltonian}
    \begin{aligned}
       &H = \underbrace{ -t \sum_{i} \left(c^{\dagger}(x_i)c(x_{i+1})+\mathrm{h.c.}\right)}_{\hat{K}}  \\
        & \underbrace{+U \sum_{i}n(x_i) n(x_{i+1})}_{\hat{U}} \; \underbrace{- \sum_{i}(v(x_i)+\mu) n(x_i)}_{\hat{V}},
    \end{aligned}
    \end{equation}
    where $c^{\dagger}(x_i)$ and $c(x_i)$ are the Fermion creation and annihilation operators acting on lattice sites $x_i$ and $n(x_i) = c^{\dagger}(x_i)c(x_i)$. $t$ is the hopping amplitude, $U$ is the density-density interaction, $v(x_i)$ is the external potential and $\mu$ is the chemical potential. Throughout this work,  we will consider the case when $t = 1$ and $U = \mu $, corresponding to an occupancy of one particle per two sites on average (half-filled band). A straightforward extension of the HK theorem to this model establishes that, on finite systems, the ground-state wave function components are a unique function of the local density $\rho_i = \langle \psi|n(x_i) |\psi\rangle$.
    
    \begin{figure*}[t]
        \includegraphics[width=1\textwidth]{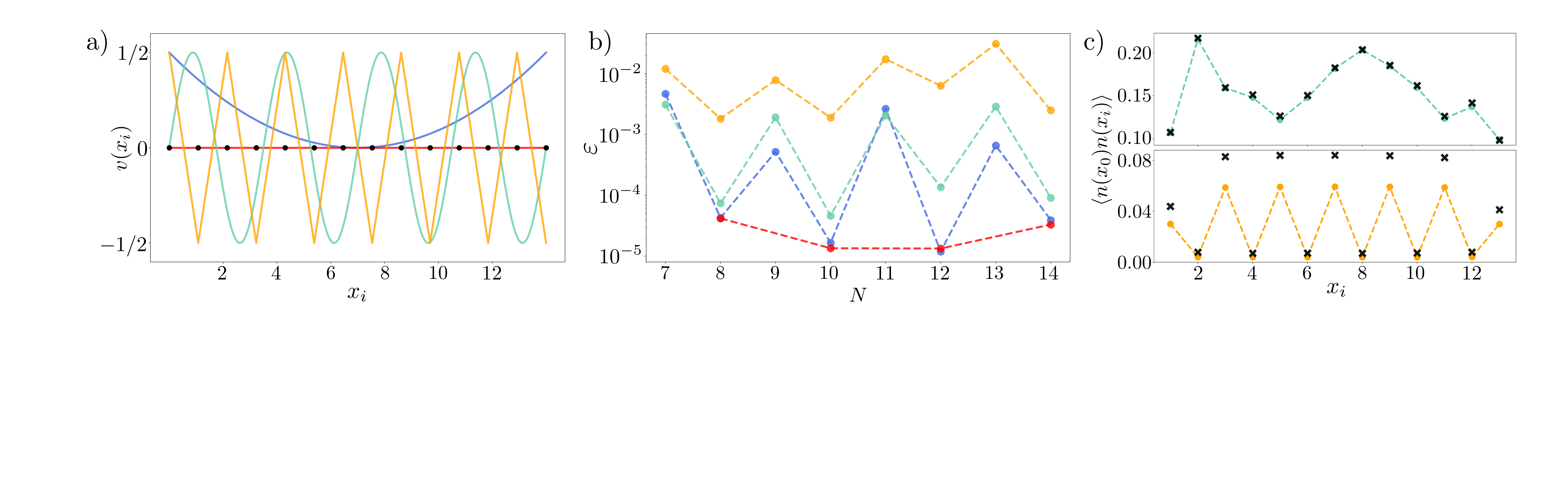}
        \caption{\label{FIG_02: Accuracy_ext_pot} Performance of the network in structured potentials. 
        \textbf{a)} Sketch of the tested potentials in a lattice with $N= 14$ sites. Potentials are quadratic (blue), no potential\footnote{Only results for an even number of lattice sites are shown in this case as the ground state with an odd number of sites is degenerate.} (red), periodic with period $N/4$ (green) and staggered (orange). Black dots represent the position of the lattice sites. Panels b) and c) 
        follow the same color code. \textbf{b)} Error as defined in Eq.~(\ref{Eq_06: Loss function}), as a function of the system size, when predicting the ground-state wave function given the potentials in panel a). \textbf{c)} Two-point density correlation functions computed from exact (dots connect by dashed lines) and ML-predicted (black crosses) wave functions. 
        }
    \end{figure*}
    
    In the absence of an external potential, the model in Eq.~(\ref{Eq_01: Full hamiltonian}) has a phase transition at $U/t = 2$ 
    of the Kosterlitz-Thoules type~\cite{Mikeska2004}. For $U/t>2$ the system is in the Mott insulator gapped phase, which is a long-range ordered charge density wave phase with spatial period $2a$, where $a$ is the lattice constant.  For $U/t<2$ the system is a gapless LL metal~\cite{Luttinger1963, Lieb1965} 
    characterized by power-law decaying correlation functions:
    \begin{equation}\label{Eq_02: correlation function LL}
        \langle n(x_i)n(x_{i+l}) \rangle \sim \frac{C_1}{l^2}+\frac{C_2(-1)^{l}}{l^{2K}},
    \end{equation}
    where $C_i$ are non-universal amplitudes and $K$ is the LL parameter, which is known exactly from Bethe ansatz: $1/K = \frac{2}{\pi}\arccos(-U/2t)$ \cite{Giamarchi2003}. The critical point follows the same scaling with an additional $\sqrt{\log(cl)}$ contribution multiplying the staggered term~\cite{Giamarchi2003, Hallberg1995}.


    \paragraph*{Density to wave function map.--}
    First, we study the possibility of learning the DTWF map using a deep fully-connected feedforward neural network \cite{Goodfellow-et-al-2016}. We consider 
    finite-size systems with $N =  7$ to $N =  14$ lattice sites. The input to the network is the $N$ values of the local density $\rho_i$ 
    and the output are the $2^N$ components of the ground-state wave function $\psi(\sigma)$ in the occupation basis: $|\psi\rangle = \sum_{\sigma}\psi(\sigma)|\sigma\rangle$, where $\sigma\equiv\{n_1,\cdots,n_N:\,n_i=0,1\}$ labels a specific occupation configuration. A representation of this architecture is shown in Figure~\ref{FIG_01: Performance} a). All the layers are connected by the composition of an affine transformation and a nonlinear rectifier function, $\textrm{Relu}(x)=\mathrm{max}(0,x)$, except for the output layer, where the chosen non linearity enforces the normalization of the wave function --see supplementary materials for further information \cite{Supplementary}--. 

        \begin{figure*}[t]
        \includegraphics[width=.95\textwidth]{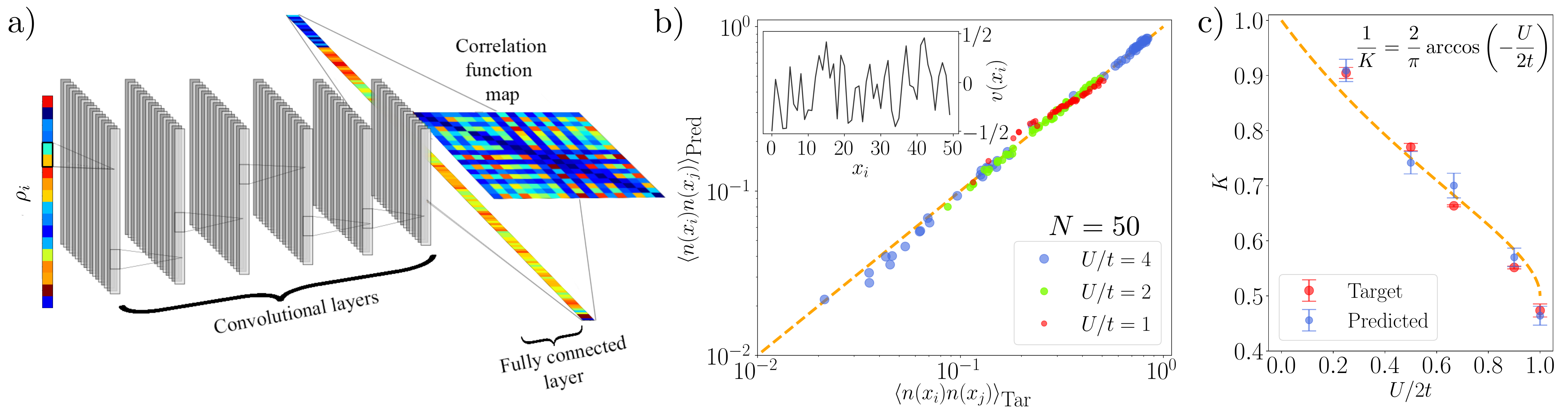}
        \caption{\label{FIG_03: ConvNet Corr functions} \textbf{a)} Scheme of the convolutional network used to represent the 
        $\rho_i \rightarrow \langle n(x_j) n(x_k) \rangle$ (DTCF) map. The output of the network is sorted to produce the correlation function map. 
        \textbf{b)} Exact versus ML-constructed values of the two-point density correlation functions given the random potential shown in the inset, for different values of $U/t$ in a lattice with $N = 50$ sites. Training sets are generated using DMRG in this case. \textbf{c)} Luttinger liquid parameter as a function of $U/2t$. The dashed line shows the exact value from Bethe ansatz. Red dots correspond to the estimated value using DMRG in a lattice with $N = 50$ sites and blue dots correspond to the estimated value from the ML-predicted correlation functions.
        } 
    \end{figure*} 
        
    Supervised learning is used to find the set parameters of the network $\{\theta\}$ that minimize the infidelity of the wave function:
        \begin{equation}\label{Eq_06: Loss function}
            \varepsilon_{\textrm{train}}(\{\theta\})
            = 1-|\langle \psi_{\textrm{tar}}|\psi_{\textrm{pred}}(\{\theta\})\rangle|,
        \end{equation}
    where $|\psi_{\textrm{tar}}\rangle$ and $|\psi_{\textrm{pred}}\rangle$ are the target and predicted wave functions respectively.
    For different $U/t$ values, different networks are trained using a set of ground-state wave functions corresponding to Hamiltonians of the form of Eq.~(\ref{Eq_01: Full hamiltonian}), with uniformly sampled random potentials with the restriction $|v(x_i)| \leq 1/2$. This ensures that the training set is not biased. The size of the training set is chosen to increase linearly with system size. The construction of the density function is tested in the two phases of the system $U/t = 1$ (metallic),  $U/t = 4$ (Mott insulator) and at the critical point $U/t = 2$. The accuracy of the learning is tested on a validation set, as per standard machine-learning practice~\cite{Goodfellow-et-al-2016}. We also test the capability of the ML-constructed wave functions to reconstruct the corresponding local densities and ground-state energies --see supplementary materials~\cite{Supplementary}--.

    Left panels in Figure~\ref{FIG_01: Performance} b), c) and d) show histograms of the error, as defined in Eq.~(\ref{Eq_06: Loss function}), 
    when constructing the wave function over the validation set for different values of $U/t$. The right panel shows correlation plots of exact versus ML-constructed wave function weights given two random potentials depicted in the insets. The histograms show narrow peaked error distributions. 
    The position of the peak is not strongly affected by the system size or the value of $U/t$. In all of the cases the peak is centered around error values 
    no larger than $\varepsilon_{\textrm{validation}} = 3.10^{-4}$ (overlaps of $|\langle \psi_{\textrm{tar}}|\psi_{\textrm{pred}}\rangle| = 0.9997$). 
    The correlation plots show that the network can accurately predict all of the wave function components, without displaying significant deviations --outside rounding errors--. Achieving a high degree of accuracy on examples not present in the training set is an indication that the network has not been over-fitted, and indeed captures the underlying connection between the local density and the ground-state wave function.

    We also test the neural network on a collection of `structured' (i.e. non-random) potentials, which are highly unlikely to belong to the training set. The potentials tested are sketched in Figure~\ref{FIG_02: Accuracy_ext_pot}~a) in a lattice with $N =14$ sites. Figure ~\ref{FIG_02: Accuracy_ext_pot} b) displays the error --as defined in Eq.~(\ref{Eq_06: Loss function})-- as a function of $N$ for the different potentials tested at the critical point $U/t = 2$. Except for the staggered potential, where the error is up to two orders of magnitude larger than the typical errors in the validation sets, all of the other cases display errors that are of the same order of  magnitude as the ones found in the validation sets. 
    
    \paragraph*{Learnability across a quantum phase transition.--}
    The lack of accuracy found in the staggered potential case is striking and points to a likely fundamental difficulty in the description of this specific regime. The quantitative and qualitative loss of accuracy in the learned DTWF map is particularly evident in two-point correlation functions, as displayed in Figure~\ref{FIG_02: Accuracy_ext_pot} c). This means that in the staggered case, the proposed ML architecture and training method are not capable of capturing the behaviour of the DTWF map in this region of the space of applied potentials.  Similar results are found in the metallic and Mott phases --see supplementary materials~\cite{Supplementary}--.
    
    To make sense of these discrepancies, we notice that the staggered potential stands on different physical grounds than the other structured cases analyzed. In this one dimensional system, an infinitesimally small staggered potential induces a QPT corresponding to the opening of a gap, in turn leading to staggered ordering of the local density ~\cite{Giamarchi2003}. More generally, this QPT is driven by the amplitude $\lambda$ of the staggered potential. In the small $\lambda$ limit, $\rho_i=\rho_s (-1)^i$ with 
    $\rho_s \propto \lambda^\mu$ the staggered density. For the $U/t$ values analyzed in this work $\mu <1$~\cite{Giamarchi2003}. Therefore, the derivative of the density with respect to the amplitude of the staggered potential diverges at the critical point. In the context of DFT, this means that the functional is not smooth in the presence of a QPT, since the density functional is constructed by a Legendre transformation from the potential, and the dependence of the density on the potential is non-analytic in this case. This non analytic behaviour leads to an intrinsic difficulty in learning the map using smooth functional approximations --such as neural networks-- and unbiased sampling from random one-body potentials. 

    \paragraph*{Density to correlation functions map.--} 
    Even in those regimes where the DTWF map is efficiently learned, the practical implementation of the statistical learning scheme is constrained to small systems by the exponential growth of the wave function with the number of lattice sites. As an alternative, here we explore the possibility of directly constructing the DTCF map with deep convolutional networks, whose size does not scale exponentially with system size. They allow us to bypass the construction of the exponentially large wave function to directly compute observables of interest, in this case density-density correlators. The input of the convolutional network is the $N$ values of $\rho_i$. The output is the $(N^2-N)/2$ different pairs of correlation functions of the system. The chosen architecture consists in six convolutional layers with $N$ filters each. The size of the kernel of the convolutions is $2$. All of the convolutional layers use a $\textrm{Relu}$ activation function. After the last convolutional layer, the output of the filters was flattened and connected using a fully connected layer with a $\textrm{Sigmoid}(x) = 1/(1+e^{-x})$ activation function. The architecture is shown in panel a) of Figure~\ref{FIG_03: ConvNet Corr functions}. Significantly larger systems are studied in this case ranging from $N = 18$ to $N = 50$ lattice sites. The training sets were the same as the ones used in the previous section, but generated using density matrix renormalization group (DMRG) on ITensor~\cite{ITensor}. The weights of the network are obtained by minimizing the relative error averaged over all the correlation pairs in each sample of the training set. Different networks are trained for the three different values of $U/t$. The performance of the networks is tested in the validation set --see supplementary materials~\cite{Supplementary}--.

    Figure~\ref{FIG_03: ConvNet Corr functions} b) shows the value of exact versus ML-constructed two-point density correlation functions in a system with $N = 50$ lattice sites, given the random potential depicted in the inset, at three different values of $U/t$. The plot demonstrates that the convolutional network is capable of accurately finding the values of the correlation functions. 
    Figure~\ref{FIG_03: ConvNet Corr functions} c) shows the value of the LL exponent --see Eq.~(\ref{Eq_02: correlation function LL})-- obtained from the DMRG and  the ML-constructed density-density correlation functions for different values of $U/t$. The value of $K$ is obtained from the power law behaviour of the correlation functions after removing the $C_1/l^2$ background. The values of $K$ obtained from the ML correlation functions are in good agreement with both the exact solution and the DMRG values
    --see supplementary materials~\cite{Supplementary} for the raw data of the correlation functions--. The small discrepancies arise from finite-size effects. Finite-size scaling allows to capture more accurate estimates of $K$ and the logarithmic correction  in the critical point --see supplementary materials~\cite{Supplementary}--. These results show that the construction of the exponentially large wave-function can be bypassed to accurately reconstruct complex observables of interest 
    and their scalings in the three regimes of (i) power law decaying correlations (ii) true long-range order (iii) critical point between the two. 
     The performance of this method is similar for $N = 18$ and $N=50$ --see supplementary materials~\cite{Supplementary}-- demonstrating its scalability.
    

    \paragraph*{Conclusion.--}In this paper we proposed a method based on supervised training deep learning to successfully construct the local density to ground-state many-body wave function (DTWF) map and the local density to correlation functions (DTCF) maps for a model of interacting spinless Fermions in a 1D lattice. Hence, providing evidence that machine learning tools provide a suitable framework to represent these high-dimensional density functionals in the two phases of the system, including the charge ordered Mott insulator, and the critical point. 
    The results serve as a proof of concept to open new lines of research where parameters from a suitable variational ansatz, such as neural network quantum states~\cite{Carleo2017}, could be predicted instead of the exponentially-many wave function components. Thanks to the insight provided by the explicit construction of the DTWF map,  we also found that the learning performance deteriorates through a quantum phase transition, due to the non analytic behaviour of the density functional. Finally, it was shown that the (exponentially costly) reconstruction of the wave function can be bypassed to directly infer complex physical observables from the local density only. An interesting open research direction concerns the application of this idea to larger system sizes. Combining our approach with an intrinsic notion of locality in the correlations and potentially transfer learning techniques could open the possibility to infer correlation functions in larger systems from learnt features in smaller ones. 

\paragraph*{Acknowledgements.--}
JRM acknowledges support from the CCQ graduate fellowship in computational quantum physics. The Flatiron Institute is a division of the Simons Foundation. The authors acknowledge discussions with Stefano Baroni, Timothy Berkelbach, Kieron Burke, Thierry Giamarchi, Masatoshi Imada, James Stokes and Miles Stoudenmire.


%


\begin{thebibliography}{44}%

\makeatletter
\providecommand \@ifxundefined [1]{%
 \@ifx{#1\undefined}
}%
\providecommand \@ifnum [1]{%
 \ifnum #1\expandafter \@firstoftwo
 \else \expandafter \@secondoftwo
 \fi
}%
\providecommand \@ifx [1]{%
 \ifx #1\expandafter \@firstoftwo
 \else \expandafter \@secondoftwo
 \fi
}%
\providecommand \natexlab [1]{#1}%
\providecommand \enquote  [1]{``#1''}%
\providecommand \bibnamefont  [1]{#1}%
\providecommand \bibfnamefont [1]{#1}%
\providecommand \citenamefont [1]{#1}%
\providecommand \href@noop [0]{\@secondoftwo}%
\providecommand \href [0]{\begingroup \@sanitize@url \@href}%
\providecommand \@href[1]{\@@startlink{#1}\@@href}%
\providecommand \@@href[1]{\endgroup#1\@@endlink}%
\providecommand \@sanitize@url [0]{\catcode `\\12\catcode `\$12\catcode
  `\&12\catcode `\#12\catcode `\^12\catcode `\_12\catcode `\%12\relax}%
\providecommand \@@startlink[1]{}%
\providecommand \@@endlink[0]{}%
\providecommand \url  [0]{\begingroup\@sanitize@url \@url }%
\providecommand \@url [1]{\endgroup\@href {#1}{\urlprefix }}%
\providecommand \urlprefix  [0]{URL }%
\providecommand \Eprint [0]{\href }%
\providecommand \doibase [0]{http://dx.doi.org/}%
\providecommand \selectlanguage [0]{\@gobble}%
\providecommand \bibinfo  [0]{\@secondoftwo}%
\providecommand \bibfield  [0]{\@secondoftwo}%
\providecommand \translation [1]{[#1]}%
\providecommand \BibitemOpen [0]{}%
\providecommand \bibitemStop [0]{}%
\providecommand \bibitemNoStop [0]{.\EOS\space}%
\providecommand \EOS [0]{\spacefactor3000\relax}%
\providecommand \BibitemShut  [1]{\csname bibitem#1\endcsname}%
\let\auto@bib@innerbib\@empty

\bibitem [{\citenamefont {Hohenberg}\ and\ \citenamefont
  {Kohn}(1964)}]{Hohenberg1964}%
  \BibitemOpen
  \bibfield  {author} {\bibinfo {author} {\bibfnamefont {P.}~\bibnamefont
  {Hohenberg}}\ and\ \bibinfo {author} {\bibfnamefont {W.}~\bibnamefont
  {Kohn}},\ }\href {\doibase 10.1103/PhysRev.136.B864} {\bibfield  {journal}
  {\bibinfo  {journal} {Phys. Rev.}\ }\textbf {\bibinfo {volume} {136}},\
  \bibinfo {pages} {B864} (\bibinfo {year} {1964})}\BibitemShut {NoStop}%
\bibitem [{\citenamefont {Cohen}\ \emph {et~al.}(2012)\citenamefont {Cohen},
  \citenamefont {Mori-Sánchez},\ and\ \citenamefont {Yang}}]{Cohen2012}%
  \BibitemOpen
  \bibfield  {author} {\bibinfo {author} {\bibfnamefont {A.~J.}\ \bibnamefont
  {Cohen}}, \bibinfo {author} {\bibfnamefont {P.}~\bibnamefont
  {Mori-Sánchez}}, \ and\ \bibinfo {author} {\bibfnamefont {W.}~\bibnamefont
  {Yang}},\ }\href {\doibase 10.1021/cr200107z} {\bibfield  {journal} {\bibinfo
   {journal} {Chem. Rev.}\ }\textbf {\bibinfo {volume} {112}},\ \bibinfo
  {pages} {289} (\bibinfo {year} {2012})}\BibitemShut {NoStop}%
\bibitem [{\citenamefont {Kohn}(1999)}]{KohnNobelLecture}%
  \BibitemOpen
  \bibfield  {author} {\bibinfo {author} {\bibfnamefont {W.}~\bibnamefont
  {Kohn}},\ }\href {\doibase 10.1103/RevModPhys.71.1253} {\bibfield  {journal}
  {\bibinfo  {journal} {Rev. Mod. Phys.}\ }\textbf {\bibinfo {volume} {71}},\
  \bibinfo {pages} {1253} (\bibinfo {year} {1999})}\BibitemShut {NoStop}%
\bibitem [{\citenamefont {Cohen}\ \emph {et~al.}(2008)\citenamefont {Cohen},
  \citenamefont {Mori-S{\'a}nchez},\ and\ \citenamefont {Yang}}]{Cohen792}%
  \BibitemOpen
  \bibfield  {author} {\bibinfo {author} {\bibfnamefont {A.~J.}\ \bibnamefont
  {Cohen}}, \bibinfo {author} {\bibfnamefont {P.}~\bibnamefont
  {Mori-S{\'a}nchez}}, \ and\ \bibinfo {author} {\bibfnamefont
  {W.}~\bibnamefont {Yang}},\ }\href {\doibase 10.1126/science.1158722}
  {\bibfield  {journal} {\bibinfo  {journal} {Science}\ }\textbf {\bibinfo
  {volume} {321}},\ \bibinfo {pages} {792} (\bibinfo {year} {2008})},\ \Eprint
  {http://arxiv.org/abs/https://science.sciencemag.org/content/321/5890/792.full.pdf}
  {https://science.sciencemag.org/content/321/5890/792.full.pdf} \BibitemShut
  {NoStop}%
\bibitem [{\citenamefont {Mori-S\'anchez}\ \emph {et~al.}(2008)\citenamefont
  {Mori-S\'anchez}, \citenamefont {Cohen},\ and\ \citenamefont
  {Yang}}]{Mori2008}%
  \BibitemOpen
  \bibfield  {author} {\bibinfo {author} {\bibfnamefont {P.}~\bibnamefont
  {Mori-S\'anchez}}, \bibinfo {author} {\bibfnamefont {A.~J.}\ \bibnamefont
  {Cohen}}, \ and\ \bibinfo {author} {\bibfnamefont {W.}~\bibnamefont {Yang}},\
  }\href {\doibase 10.1103/PhysRevLett.100.146401} {\bibfield  {journal}
  {\bibinfo  {journal} {Phys. Rev. Lett.}\ }\textbf {\bibinfo {volume} {100}},\
  \bibinfo {pages} {146401} (\bibinfo {year} {2008})}\BibitemShut {NoStop}%
\bibitem [{\citenamefont {Carleo}\ \emph {et~al.}(2019)\citenamefont {Carleo},
  \citenamefont {Cirac}, \citenamefont {Cranmer}, \citenamefont {Daudet},
  \citenamefont {Schuld}, \citenamefont {Tishby}, \citenamefont
  {Vogt-Maranto},\ and\ \citenamefont {Zdeborová}}]{Carleo2019}%
  \BibitemOpen
  \bibfield  {author} {\bibinfo {author} {\bibfnamefont {G.}~\bibnamefont
  {Carleo}}, \bibinfo {author} {\bibfnamefont {I.}~\bibnamefont {Cirac}},
  \bibinfo {author} {\bibfnamefont {K.}~\bibnamefont {Cranmer}}, \bibinfo
  {author} {\bibfnamefont {L.}~\bibnamefont {Daudet}}, \bibinfo {author}
  {\bibfnamefont {M.}~\bibnamefont {Schuld}}, \bibinfo {author} {\bibfnamefont
  {N.}~\bibnamefont {Tishby}}, \bibinfo {author} {\bibfnamefont
  {L.}~\bibnamefont {Vogt-Maranto}}, \ and\ \bibinfo {author} {\bibfnamefont
  {L.}~\bibnamefont {Zdeborová}},\ }\href {\doibase
  10.1103/RevModPhys.91.045002} {\bibfield  {journal} {\bibinfo  {journal}
  {Reviews of Modern Physics}\ }\textbf {\bibinfo {volume} {91}},\ \bibinfo
  {pages} {045002} (\bibinfo {year} {2019})}\BibitemShut {NoStop}%
\bibitem [{\citenamefont {Kohn}\ and\ \citenamefont {Sham}(1965)}]{Kohn1965}%
  \BibitemOpen
  \bibfield  {author} {\bibinfo {author} {\bibfnamefont {W.}~\bibnamefont
  {Kohn}}\ and\ \bibinfo {author} {\bibfnamefont {L.~J.}\ \bibnamefont
  {Sham}},\ }\href {\doibase 10.1103/PhysRev.140.A1133} {\bibfield  {journal}
  {\bibinfo  {journal} {Phys. Rev.}\ }\textbf {\bibinfo {volume} {140}},\
  \bibinfo {pages} {A1133} (\bibinfo {year} {1965})}\BibitemShut {NoStop}%
\bibitem [{\citenamefont {Kolb}\ \emph {et~al.}(2017)\citenamefont {Kolb},
  \citenamefont {Lentz},\ and\ \citenamefont {Kolpak}}]{Kolb2017}%
  \BibitemOpen
  \bibfield  {author} {\bibinfo {author} {\bibfnamefont {B.}~\bibnamefont
  {Kolb}}, \bibinfo {author} {\bibfnamefont {L.~C.}\ \bibnamefont {Lentz}}, \
  and\ \bibinfo {author} {\bibfnamefont {A.~M.}\ \bibnamefont {Kolpak}},\
  }\href {\doibase 10.1038/s41598-017-01251-z} {\bibfield  {journal} {\bibinfo
  {journal} {Scientific Reports}\ }\textbf {\bibinfo {volume} {7}},\ \bibinfo
  {pages} {1192} (\bibinfo {year} {2017})}\BibitemShut {NoStop}%
\bibitem [{\citenamefont {Liu}\ \emph {et~al.}(2017{\natexlab{a}})\citenamefont
  {Liu}, \citenamefont {Wang}, \citenamefont {Du}, \citenamefont {Hu},
  \citenamefont {Zheng},\ and\ \citenamefont {Chen}}]{Liu2017}%
  \BibitemOpen
  \bibfield  {author} {\bibinfo {author} {\bibfnamefont {Q.}~\bibnamefont
  {Liu}}, \bibinfo {author} {\bibfnamefont {J.}~\bibnamefont {Wang}}, \bibinfo
  {author} {\bibfnamefont {P.}~\bibnamefont {Du}}, \bibinfo {author}
  {\bibfnamefont {L.}~\bibnamefont {Hu}}, \bibinfo {author} {\bibfnamefont
  {X.}~\bibnamefont {Zheng}}, \ and\ \bibinfo {author} {\bibfnamefont
  {G.}~\bibnamefont {Chen}},\ }\href {\doibase 10.1021/acs.jpca.7b07045}
  {\bibfield  {journal} {\bibinfo  {journal} {The Journal of Physical Chemistry
  A}\ }\textbf {\bibinfo {volume} {121}},\ \bibinfo {pages} {7273} (\bibinfo
  {year} {2017}{\natexlab{a}})},\ \bibinfo {note} {pMID: 28876064},\ \Eprint
  {http://arxiv.org/abs/https://doi.org/10.1021/acs.jpca.7b07045}
  {https://doi.org/10.1021/acs.jpca.7b07045} \BibitemShut {NoStop}%
\bibitem [{\citenamefont {Lundgaard}\ \emph {et~al.}(2016)\citenamefont
  {Lundgaard}, \citenamefont {Wellendorff}, \citenamefont {Voss}, \citenamefont
  {Jacobsen},\ and\ \citenamefont {Bligaard}}]{Lundgaard2016}%
  \BibitemOpen
  \bibfield  {author} {\bibinfo {author} {\bibfnamefont {K.~T.}\ \bibnamefont
  {Lundgaard}}, \bibinfo {author} {\bibfnamefont {J.}~\bibnamefont
  {Wellendorff}}, \bibinfo {author} {\bibfnamefont {J.}~\bibnamefont {Voss}},
  \bibinfo {author} {\bibfnamefont {K.~W.}\ \bibnamefont {Jacobsen}}, \ and\
  \bibinfo {author} {\bibfnamefont {T.}~\bibnamefont {Bligaard}},\ }\href
  {\doibase 10.1103/PhysRevB.93.235162} {\bibfield  {journal} {\bibinfo
  {journal} {Phys. Rev. B}\ }\textbf {\bibinfo {volume} {93}},\ \bibinfo
  {pages} {235162} (\bibinfo {year} {2016})}\BibitemShut {NoStop}%
\bibitem [{\citenamefont {Pilati}\ and\ \citenamefont
  {Pieri}(2019)}]{Nudejima2019}%
  \BibitemOpen
  \bibfield  {author} {\bibinfo {author} {\bibfnamefont {S.}~\bibnamefont
  {Pilati}}\ and\ \bibinfo {author} {\bibfnamefont {P.}~\bibnamefont {Pieri}},\
  }\href {\doibase 10.1063/1.5100165} {\bibfield  {journal} {\bibinfo
  {journal} {Scientific Reports}\ }\textbf {\bibinfo {volume} {9}},\ \bibinfo
  {pages} {5613} (\bibinfo {year} {2019})},\ \Eprint
  {http://arxiv.org/abs/https://doi.org/10.1063/1.5100165}
  {https://doi.org/10.1063/1.5100165} \BibitemShut {NoStop}%
\bibitem [{\citenamefont {{Schmidt}}\ \emph {et~al.}(2019)\citenamefont
  {{Schmidt}}, \citenamefont {{Benavides-Riveros}},\ and\ \citenamefont
  {{Marques}}}]{Schmidt2019}%
  \BibitemOpen
  \bibfield  {author} {\bibinfo {author} {\bibfnamefont {J.}~\bibnamefont
  {{Schmidt}}}, \bibinfo {author} {\bibfnamefont {C.~L.}\ \bibnamefont
  {{Benavides-Riveros}}}, \ and\ \bibinfo {author} {\bibfnamefont {M.~A.~L.}\
  \bibnamefont {{Marques}}},\ }\href@noop {} {\bibfield  {journal} {\bibinfo
  {journal} {arXiv e-prints}\ ,\ \bibinfo {eid} {arXiv:1908.06198}} (\bibinfo
  {year} {2019})},\ \Eprint {http://arxiv.org/abs/1908.06198} {arXiv:1908.06198
  [physics.comp-ph]} \BibitemShut {NoStop}%
\bibitem [{\citenamefont {Nagai}\ \emph {et~al.}(2018)\citenamefont {Nagai},
  \citenamefont {Akashi}, \citenamefont {Sasaki},\ and\ \citenamefont
  {Tsuneyuki}}]{Nagai2018}%
  \BibitemOpen
  \bibfield  {author} {\bibinfo {author} {\bibfnamefont {R.}~\bibnamefont
  {Nagai}}, \bibinfo {author} {\bibfnamefont {R.}~\bibnamefont {Akashi}},
  \bibinfo {author} {\bibfnamefont {S.}~\bibnamefont {Sasaki}}, \ and\ \bibinfo
  {author} {\bibfnamefont {S.}~\bibnamefont {Tsuneyuki}},\ }\href {\doibase
  10.1063/1.5029279} {\bibfield  {journal} {\bibinfo  {journal} {The Journal of
  Chemical Physics}\ }\textbf {\bibinfo {volume} {148}},\ \bibinfo {pages}
  {241737} (\bibinfo {year} {2018})},\ \Eprint
  {http://arxiv.org/abs/https://doi.org/10.1063/1.5029279}
  {https://doi.org/10.1063/1.5029279} \BibitemShut {NoStop}%
\bibitem [{\citenamefont {Dick}\ and\ \citenamefont
  {Fernandez-Serra}(2019)}]{Dick2019}%
  \BibitemOpen
  \bibfield  {author} {\bibinfo {author} {\bibfnamefont {S.}~\bibnamefont
  {Dick}}\ and\ \bibinfo {author} {\bibfnamefont {M.}~\bibnamefont
  {Fernandez-Serra}},\ }\href {\doibase 10.26434/chemrxiv.9947312.v1} {\
  (\bibinfo {year} {2019}),\ 10.26434/chemrxiv.9947312.v1}\BibitemShut
  {NoStop}%
\bibitem [{\citenamefont {Golub}\ and\ \citenamefont
  {Manzhos}(2019)}]{Golub2019}%
  \BibitemOpen
  \bibfield  {author} {\bibinfo {author} {\bibfnamefont {P.}~\bibnamefont
  {Golub}}\ and\ \bibinfo {author} {\bibfnamefont {S.}~\bibnamefont
  {Manzhos}},\ }\href {\doibase 10.1039/C8CP06433D} {\bibfield  {journal}
  {\bibinfo  {journal} {Phys. Chem. Chem. Phys.}\ }\textbf {\bibinfo {volume}
  {21}},\ \bibinfo {pages} {378} (\bibinfo {year} {2019})}\BibitemShut
  {NoStop}%
\bibitem [{\citenamefont {Seino}\ \emph {et~al.}(2018)\citenamefont {Seino},
  \citenamefont {Kageyama}, \citenamefont {Fujinami}, \citenamefont {Ikabata},\
  and\ \citenamefont {Nakai}}]{Seino2018}%
  \BibitemOpen
  \bibfield  {author} {\bibinfo {author} {\bibfnamefont {J.}~\bibnamefont
  {Seino}}, \bibinfo {author} {\bibfnamefont {R.}~\bibnamefont {Kageyama}},
  \bibinfo {author} {\bibfnamefont {M.}~\bibnamefont {Fujinami}}, \bibinfo
  {author} {\bibfnamefont {Y.}~\bibnamefont {Ikabata}}, \ and\ \bibinfo
  {author} {\bibfnamefont {H.}~\bibnamefont {Nakai}},\ }\href {\doibase
  10.1063/1.5007230} {\bibfield  {journal} {\bibinfo  {journal} {The Journal of
  Chemical Physics}\ }\textbf {\bibinfo {volume} {148}},\ \bibinfo {pages}
  {241705} (\bibinfo {year} {2018})},\ \Eprint
  {http://arxiv.org/abs/https://doi.org/10.1063/1.5007230}
  {https://doi.org/10.1063/1.5007230} \BibitemShut {NoStop}%
\bibitem [{\citenamefont {Snyder}\ \emph {et~al.}(2013)\citenamefont {Snyder},
  \citenamefont {Rupp}, \citenamefont {Hansen}, \citenamefont {Blooston},
  \citenamefont {Müller},\ and\ \citenamefont {Burke}}]{Snyder2013}%
  \BibitemOpen
  \bibfield  {author} {\bibinfo {author} {\bibfnamefont {J.~C.}\ \bibnamefont
  {Snyder}}, \bibinfo {author} {\bibfnamefont {M.}~\bibnamefont {Rupp}},
  \bibinfo {author} {\bibfnamefont {K.}~\bibnamefont {Hansen}}, \bibinfo
  {author} {\bibfnamefont {L.}~\bibnamefont {Blooston}}, \bibinfo {author}
  {\bibfnamefont {K.-R.}\ \bibnamefont {Müller}}, \ and\ \bibinfo {author}
  {\bibfnamefont {K.}~\bibnamefont {Burke}},\ }\href {\doibase
  10.1063/1.4834075} {\bibfield  {journal} {\bibinfo  {journal} {The Journal of
  Chemical Physics}\ }\textbf {\bibinfo {volume} {139}},\ \bibinfo {pages}
  {224104} (\bibinfo {year} {2013})},\ \Eprint
  {http://arxiv.org/abs/https://doi.org/10.1063/1.4834075}
  {https://doi.org/10.1063/1.4834075} \BibitemShut {NoStop}%
\bibitem [{\citenamefont {Li}\ \emph {et~al.}(2016{\natexlab{a}})\citenamefont
  {Li}, \citenamefont {Snyder}, \citenamefont {Pelaschier}, \citenamefont
  {Huang}, \citenamefont {Niranjan}, \citenamefont {Duncan}, \citenamefont
  {Rupp}, \citenamefont {Müller},\ and\ \citenamefont {Burke}}]{Snyder2016}%
  \BibitemOpen
  \bibfield  {author} {\bibinfo {author} {\bibfnamefont {L.}~\bibnamefont
  {Li}}, \bibinfo {author} {\bibfnamefont {J.~C.}\ \bibnamefont {Snyder}},
  \bibinfo {author} {\bibfnamefont {I.~M.}\ \bibnamefont {Pelaschier}},
  \bibinfo {author} {\bibfnamefont {J.}~\bibnamefont {Huang}}, \bibinfo
  {author} {\bibfnamefont {U.-N.}\ \bibnamefont {Niranjan}}, \bibinfo {author}
  {\bibfnamefont {P.}~\bibnamefont {Duncan}}, \bibinfo {author} {\bibfnamefont
  {M.}~\bibnamefont {Rupp}}, \bibinfo {author} {\bibfnamefont {K.-R.}\
  \bibnamefont {Müller}}, \ and\ \bibinfo {author} {\bibfnamefont
  {K.}~\bibnamefont {Burke}},\ }\href {\doibase 10.1002/qua.25040} {\bibfield
  {journal} {\bibinfo  {journal} {International Journal of Quantum Chemistry}\
  }\textbf {\bibinfo {volume} {116}},\ \bibinfo {pages} {819} (\bibinfo {year}
  {2016}{\natexlab{a}})},\ \Eprint
  {http://arxiv.org/abs/https://onlinelibrary.wiley.com/doi/pdf/10.1002/qua.25040}
  {https://onlinelibrary.wiley.com/doi/pdf/10.1002/qua.25040} \BibitemShut
  {NoStop}%
\bibitem [{\citenamefont {Snyder}\ \emph {et~al.}(2012)\citenamefont {Snyder},
  \citenamefont {Rupp}, \citenamefont {Hansen}, \citenamefont {M\"uller},\ and\
  \citenamefont {Burke}}]{Snyder2012}%
  \BibitemOpen
  \bibfield  {author} {\bibinfo {author} {\bibfnamefont {J.~C.}\ \bibnamefont
  {Snyder}}, \bibinfo {author} {\bibfnamefont {M.}~\bibnamefont {Rupp}},
  \bibinfo {author} {\bibfnamefont {K.}~\bibnamefont {Hansen}}, \bibinfo
  {author} {\bibfnamefont {K.-R.}\ \bibnamefont {M\"uller}}, \ and\ \bibinfo
  {author} {\bibfnamefont {K.}~\bibnamefont {Burke}},\ }\href {\doibase
  10.1103/PhysRevLett.108.253002} {\bibfield  {journal} {\bibinfo  {journal}
  {Phys. Rev. Lett.}\ }\textbf {\bibinfo {volume} {108}},\ \bibinfo {pages}
  {253002} (\bibinfo {year} {2012})}\BibitemShut {NoStop}%
\bibitem [{\citenamefont {Li}\ \emph {et~al.}(2016{\natexlab{b}})\citenamefont
  {Li}, \citenamefont {Baker}, \citenamefont {White},\ and\ \citenamefont
  {Burke}}]{Li2016}%
  \BibitemOpen
  \bibfield  {author} {\bibinfo {author} {\bibfnamefont {L.}~\bibnamefont
  {Li}}, \bibinfo {author} {\bibfnamefont {T.~E.}\ \bibnamefont {Baker}},
  \bibinfo {author} {\bibfnamefont {S.~R.}\ \bibnamefont {White}}, \ and\
  \bibinfo {author} {\bibfnamefont {K.}~\bibnamefont {Burke}},\ }\href
  {\doibase 10.1103/PhysRevB.94.245129} {\bibfield  {journal} {\bibinfo
  {journal} {Phys. Rev. B}\ }\textbf {\bibinfo {volume} {94}},\ \bibinfo
  {pages} {245129} (\bibinfo {year} {2016}{\natexlab{b}})}\BibitemShut
  {NoStop}%
\bibitem [{\citenamefont {Brockherde}\ \emph {et~al.}(2017)\citenamefont
  {Brockherde}, \citenamefont {Vogt}, \citenamefont {Li}, \citenamefont
  {Tuckerman}, \citenamefont {Burke},\ and\ \citenamefont
  {Müller}}]{Brockherde2017}%
  \BibitemOpen
  \bibfield  {author} {\bibinfo {author} {\bibfnamefont {F.}~\bibnamefont
  {Brockherde}}, \bibinfo {author} {\bibfnamefont {L.}~\bibnamefont {Vogt}},
  \bibinfo {author} {\bibfnamefont {L.}~\bibnamefont {Li}}, \bibinfo {author}
  {\bibfnamefont {M.~E.}\ \bibnamefont {Tuckerman}}, \bibinfo {author}
  {\bibfnamefont {K.}~\bibnamefont {Burke}}, \ and\ \bibinfo {author}
  {\bibfnamefont {K.-R.}\ \bibnamefont {Müller}},\ }\href
  {https://doi.org/10.1038/s41467-017-00839-3} {\bibfield  {journal} {\bibinfo
  {journal} {Nature Communications}\ }\textbf {\bibinfo {volume} {8}},\
  \bibinfo {pages} {872} (\bibinfo {year} {2017})}\BibitemShut {NoStop}%
\bibitem [{\citenamefont {Liu}\ \emph {et~al.}(2017{\natexlab{b}})\citenamefont
  {Liu}, \citenamefont {Wang}, \citenamefont {Du}, \citenamefont {Hu},
  \citenamefont {Zheng},\ and\ \citenamefont {Chen}}]{Pilati2019}%
  \BibitemOpen
  \bibfield  {author} {\bibinfo {author} {\bibfnamefont {Q.}~\bibnamefont
  {Liu}}, \bibinfo {author} {\bibfnamefont {J.}~\bibnamefont {Wang}}, \bibinfo
  {author} {\bibfnamefont {P.}~\bibnamefont {Du}}, \bibinfo {author}
  {\bibfnamefont {L.}~\bibnamefont {Hu}}, \bibinfo {author} {\bibfnamefont
  {X.}~\bibnamefont {Zheng}}, \ and\ \bibinfo {author} {\bibfnamefont
  {G.}~\bibnamefont {Chen}},\ }\href {\doibase 10.1038/s41598-019-42125-w}
  {\bibfield  {journal} {\bibinfo  {journal} {The Journal of Physical Chemistry
  A}\ }\textbf {\bibinfo {volume} {121}},\ \bibinfo {pages} {7273} (\bibinfo
  {year} {2017}{\natexlab{b}})},\ \Eprint
  {http://arxiv.org/abs/https://doi.org/10.1038/s41598-019-42125-w}
  {https://doi.org/10.1038/s41598-019-42125-w} \BibitemShut {NoStop}%
\bibitem [{\citenamefont {Nelson}\ \emph {et~al.}(2019)\citenamefont {Nelson},
  \citenamefont {Tiwari},\ and\ \citenamefont {Sanvito}}]{Nelson2019}%
  \BibitemOpen
  \bibfield  {author} {\bibinfo {author} {\bibfnamefont {J.}~\bibnamefont
  {Nelson}}, \bibinfo {author} {\bibfnamefont {R.}~\bibnamefont {Tiwari}}, \
  and\ \bibinfo {author} {\bibfnamefont {S.}~\bibnamefont {Sanvito}},\ }\href
  {\doibase 10.1103/PhysRevB.99.075132} {\bibfield  {journal} {\bibinfo
  {journal} {Phys. Rev. B}\ }\textbf {\bibinfo {volume} {99}},\ \bibinfo
  {pages} {075132} (\bibinfo {year} {2019})}\BibitemShut {NoStop}%
\bibitem [{\citenamefont {Ryczko}\ \emph {et~al.}(2019)\citenamefont {Ryczko},
  \citenamefont {Strubbe},\ and\ \citenamefont {Tamblyn}}]{Ryczko2019}%
  \BibitemOpen
  \bibfield  {author} {\bibinfo {author} {\bibfnamefont {K.}~\bibnamefont
  {Ryczko}}, \bibinfo {author} {\bibfnamefont {D.~A.}\ \bibnamefont {Strubbe}},
  \ and\ \bibinfo {author} {\bibfnamefont {I.}~\bibnamefont {Tamblyn}},\ }\href
  {\doibase 10.1103/PhysRevA.100.022512} {\bibfield  {journal} {\bibinfo
  {journal} {Phys. Rev. A}\ }\textbf {\bibinfo {volume} {100}},\ \bibinfo
  {pages} {022512} (\bibinfo {year} {2019})}\BibitemShut {NoStop}%
\bibitem [{\citenamefont {Grabowsky}\ \emph {et~al.}(2017)\citenamefont
  {Grabowsky}, \citenamefont {Genoni},\ and\ \citenamefont
  {Bürgi}}]{Grabowsky2017}%
  \BibitemOpen
  \bibfield  {author} {\bibinfo {author} {\bibfnamefont {S.}~\bibnamefont
  {Grabowsky}}, \bibinfo {author} {\bibfnamefont {A.}~\bibnamefont {Genoni}}, \
  and\ \bibinfo {author} {\bibfnamefont {H.-B.}\ \bibnamefont {Bürgi}},\
  }\href {\doibase 10.1039/C6SC05504D} {\bibfield  {journal} {\bibinfo
  {journal} {Chem. Sci.}\ }\textbf {\bibinfo {volume} {8}},\ \bibinfo {pages}
  {4159} (\bibinfo {year} {2017})}\BibitemShut {NoStop}%
\bibitem [{\citenamefont {Tolborg}\ and\ \citenamefont
  {Iversen}(2019)}]{Tolborg2019}%
  \BibitemOpen
  \bibfield  {author} {\bibinfo {author} {\bibfnamefont {K.}~\bibnamefont
  {Tolborg}}\ and\ \bibinfo {author} {\bibfnamefont {B.~B.}\ \bibnamefont
  {Iversen}},\ }\href {\doibase 10.1002/chem.201903087} {\bibfield  {journal}
  {\bibinfo  {journal} {Chemistry – A European Journal}\ }\textbf {\bibinfo
  {volume} {0}} (\bibinfo {year} {2019}),\ 10.1002/chem.201903087},\ \Eprint
  {http://arxiv.org/abs/https://onlinelibrary.wiley.com/doi/pdf/10.1002/chem.201903087}
  {https://onlinelibrary.wiley.com/doi/pdf/10.1002/chem.201903087} \BibitemShut
  {NoStop}%
\bibitem [{\citenamefont {Lye}\ \emph {et~al.}(2005)\citenamefont {Lye},
  \citenamefont {Fallani}, \citenamefont {Modugno}, \citenamefont {Wiersma},
  \citenamefont {Fort},\ and\ \citenamefont {Inguscio}}]{Lye2005}%
  \BibitemOpen
  \bibfield  {author} {\bibinfo {author} {\bibfnamefont {J.~E.}\ \bibnamefont
  {Lye}}, \bibinfo {author} {\bibfnamefont {L.}~\bibnamefont {Fallani}},
  \bibinfo {author} {\bibfnamefont {M.}~\bibnamefont {Modugno}}, \bibinfo
  {author} {\bibfnamefont {D.~S.}\ \bibnamefont {Wiersma}}, \bibinfo {author}
  {\bibfnamefont {C.}~\bibnamefont {Fort}}, \ and\ \bibinfo {author}
  {\bibfnamefont {M.}~\bibnamefont {Inguscio}},\ }\href {\doibase
  10.1103/PhysRevLett.95.070401} {\bibfield  {journal} {\bibinfo  {journal}
  {Phys. Rev. Lett.}\ }\textbf {\bibinfo {volume} {95}},\ \bibinfo {pages}
  {070401} (\bibinfo {year} {2005})}\BibitemShut {NoStop}%
\bibitem [{\citenamefont {White}\ \emph {et~al.}(2009)\citenamefont {White},
  \citenamefont {Pasienski}, \citenamefont {McKay}, \citenamefont {Zhou},
  \citenamefont {Ceperley},\ and\ \citenamefont {DeMarco}}]{White2009}%
  \BibitemOpen
  \bibfield  {author} {\bibinfo {author} {\bibfnamefont {M.}~\bibnamefont
  {White}}, \bibinfo {author} {\bibfnamefont {M.}~\bibnamefont {Pasienski}},
  \bibinfo {author} {\bibfnamefont {D.}~\bibnamefont {McKay}}, \bibinfo
  {author} {\bibfnamefont {S.~Q.}\ \bibnamefont {Zhou}}, \bibinfo {author}
  {\bibfnamefont {D.}~\bibnamefont {Ceperley}}, \ and\ \bibinfo {author}
  {\bibfnamefont {B.}~\bibnamefont {DeMarco}},\ }\href {\doibase
  10.1103/PhysRevLett.102.055301} {\bibfield  {journal} {\bibinfo  {journal}
  {Phys. Rev. Lett.}\ }\textbf {\bibinfo {volume} {102}},\ \bibinfo {pages}
  {055301} (\bibinfo {year} {2009})}\BibitemShut {NoStop}%
\bibitem [{\citenamefont {Celi}\ \emph {et~al.}(2019)\citenamefont {Celi},
  \citenamefont {Vermersch}, \citenamefont {Viyuela}, \citenamefont {Pichler},
  \citenamefont {Lukin},\ and\ \citenamefont {Zoller}}]{Celi2019}%
  \BibitemOpen
  \bibfield  {author} {\bibinfo {author} {\bibfnamefont {A.}~\bibnamefont
  {Celi}}, \bibinfo {author} {\bibfnamefont {B.}~\bibnamefont {Vermersch}},
  \bibinfo {author} {\bibfnamefont {O.}~\bibnamefont {Viyuela}}, \bibinfo
  {author} {\bibfnamefont {H.}~\bibnamefont {Pichler}}, \bibinfo {author}
  {\bibfnamefont {M.~D.}\ \bibnamefont {Lukin}}, \ and\ \bibinfo {author}
  {\bibfnamefont {P.}~\bibnamefont {Zoller}},\ }\href@noop {} {\  (\bibinfo
  {year} {2019})},\ \Eprint {http://arxiv.org/abs/1907.03311} {arXiv:1907.03311
  [quant-ph]} \BibitemShut {NoStop}%
\bibitem [{\citenamefont {Omran}\ \emph {et~al.}(2019)\citenamefont {Omran},
  \citenamefont {Levine}, \citenamefont {Keesling}, \citenamefont {Semeghini},
  \citenamefont {Wang}, \citenamefont {Ebadi}, \citenamefont {Bernien},
  \citenamefont {Zibrov}, \citenamefont {Pichler}, \citenamefont {Choi},\ and\
  \citenamefont {et~al.}}]{Omran2019}%
  \BibitemOpen
  \bibfield  {author} {\bibinfo {author} {\bibfnamefont {A.}~\bibnamefont
  {Omran}}, \bibinfo {author} {\bibfnamefont {H.}~\bibnamefont {Levine}},
  \bibinfo {author} {\bibfnamefont {A.}~\bibnamefont {Keesling}}, \bibinfo
  {author} {\bibfnamefont {G.}~\bibnamefont {Semeghini}}, \bibinfo {author}
  {\bibfnamefont {T.~T.}\ \bibnamefont {Wang}}, \bibinfo {author}
  {\bibfnamefont {S.}~\bibnamefont {Ebadi}}, \bibinfo {author} {\bibfnamefont
  {H.}~\bibnamefont {Bernien}}, \bibinfo {author} {\bibfnamefont {A.~S.}\
  \bibnamefont {Zibrov}}, \bibinfo {author} {\bibfnamefont {H.}~\bibnamefont
  {Pichler}}, \bibinfo {author} {\bibfnamefont {S.}~\bibnamefont {Choi}}, \
  and\ \bibinfo {author} {\bibnamefont {et~al.}},\ }\href {\doibase
  10.1126/science.aax9743} {\bibfield  {journal} {\bibinfo  {journal}
  {Science}\ }\textbf {\bibinfo {volume} {365}},\ \bibinfo {pages} {570–574}
  (\bibinfo {year} {2019})}\BibitemShut {NoStop}%
\bibitem [{\citenamefont {Keesling}\ \emph {et~al.}(2019)\citenamefont
  {Keesling}, \citenamefont {Omran}, \citenamefont {Levine}, \citenamefont
  {Bernien}, \citenamefont {Pichler}, \citenamefont {Choi}, \citenamefont
  {Samajdar}, \citenamefont {Schwartz}, \citenamefont {Silvi}, \citenamefont
  {Sachdev}, \citenamefont {Zoller}, \citenamefont {Endres}, \citenamefont
  {Greiner}, \citenamefont {Vuletic},\ and\ \citenamefont
  {Lukin}}]{Keesling2019}%
  \BibitemOpen
  \bibfield  {author} {\bibinfo {author} {\bibfnamefont {A.}~\bibnamefont
  {Keesling}}, \bibinfo {author} {\bibfnamefont {A.}~\bibnamefont {Omran}},
  \bibinfo {author} {\bibfnamefont {H.}~\bibnamefont {Levine}}, \bibinfo
  {author} {\bibfnamefont {H.}~\bibnamefont {Bernien}}, \bibinfo {author}
  {\bibfnamefont {H.}~\bibnamefont {Pichler}}, \bibinfo {author} {\bibfnamefont
  {S.}~\bibnamefont {Choi}}, \bibinfo {author} {\bibfnamefont {R.}~\bibnamefont
  {Samajdar}}, \bibinfo {author} {\bibfnamefont {S.}~\bibnamefont {Schwartz}},
  \bibinfo {author} {\bibfnamefont {P.}~\bibnamefont {Silvi}}, \bibinfo
  {author} {\bibfnamefont {S.}~\bibnamefont {Sachdev}}, \bibinfo {author}
  {\bibfnamefont {P.}~\bibnamefont {Zoller}}, \bibinfo {author} {\bibfnamefont
  {M.}~\bibnamefont {Endres}}, \bibinfo {author} {\bibfnamefont
  {M.}~\bibnamefont {Greiner}}, \bibinfo {author} {\bibfnamefont
  {V.}~\bibnamefont {Vuletic}}, \ and\ \bibinfo {author} {\bibfnamefont
  {M.~D.}\ \bibnamefont {Lukin}},\ }\href {\doibase 10.1038/s41586-019-1070-1}
  {\bibfield  {journal} {\bibinfo  {journal} {Nature}\ }\textbf {\bibinfo
  {volume} {568}},\ \bibinfo {pages} {207} (\bibinfo {year}
  {2019})}\BibitemShut {NoStop}%
\bibitem [{\citenamefont {Bloch}\ \emph {et~al.}(2008)\citenamefont {Bloch},
  \citenamefont {Dalibard},\ and\ \citenamefont {Zwerger}}]{Bloch2008}%
  \BibitemOpen
  \bibfield  {author} {\bibinfo {author} {\bibfnamefont {I.}~\bibnamefont
  {Bloch}}, \bibinfo {author} {\bibfnamefont {J.}~\bibnamefont {Dalibard}}, \
  and\ \bibinfo {author} {\bibfnamefont {W.}~\bibnamefont {Zwerger}},\ }\href
  {\doibase 10.1103/RevModPhys.80.885} {\bibfield  {journal} {\bibinfo
  {journal} {Rev. Mod. Phys.}\ }\textbf {\bibinfo {volume} {80}},\ \bibinfo
  {pages} {885} (\bibinfo {year} {2008})}\BibitemShut {NoStop}%
\bibitem [{\citenamefont {Scherg}\ \emph {et~al.}(2018)\citenamefont {Scherg},
  \citenamefont {Kohlert}, \citenamefont {Herbrych}, \citenamefont {Stolpp},
  \citenamefont {Bordia}, \citenamefont {Schneider}, \citenamefont
  {Heidrich-Meisner}, \citenamefont {Bloch},\ and\ \citenamefont
  {Aidelsburger}}]{Scherg2018}%
  \BibitemOpen
  \bibfield  {author} {\bibinfo {author} {\bibfnamefont {S.}~\bibnamefont
  {Scherg}}, \bibinfo {author} {\bibfnamefont {T.}~\bibnamefont {Kohlert}},
  \bibinfo {author} {\bibfnamefont {J.}~\bibnamefont {Herbrych}}, \bibinfo
  {author} {\bibfnamefont {J.}~\bibnamefont {Stolpp}}, \bibinfo {author}
  {\bibfnamefont {P.}~\bibnamefont {Bordia}}, \bibinfo {author} {\bibfnamefont
  {U.}~\bibnamefont {Schneider}}, \bibinfo {author} {\bibfnamefont
  {F.}~\bibnamefont {Heidrich-Meisner}}, \bibinfo {author} {\bibfnamefont
  {I.}~\bibnamefont {Bloch}}, \ and\ \bibinfo {author} {\bibfnamefont
  {M.}~\bibnamefont {Aidelsburger}},\ }\href {\doibase
  10.1103/PhysRevLett.121.130402} {\bibfield  {journal} {\bibinfo  {journal}
  {Phys. Rev. Lett.}\ }\textbf {\bibinfo {volume} {121}},\ \bibinfo {pages}
  {130402} (\bibinfo {year} {2018})}\BibitemShut {NoStop}%
\bibitem [{\citenamefont {Sherson}\ \emph {et~al.}(2010)\citenamefont
  {Sherson}, \citenamefont {Weitenberg}, \citenamefont {Endres}, \citenamefont
  {Cheneau}, \citenamefont {Bloch},\ and\ \citenamefont {Kuhr}}]{Sherson2010}%
  \BibitemOpen
  \bibfield  {author} {\bibinfo {author} {\bibfnamefont {J.~F.}\ \bibnamefont
  {Sherson}}, \bibinfo {author} {\bibfnamefont {C.}~\bibnamefont {Weitenberg}},
  \bibinfo {author} {\bibfnamefont {M.}~\bibnamefont {Endres}}, \bibinfo
  {author} {\bibfnamefont {M.}~\bibnamefont {Cheneau}}, \bibinfo {author}
  {\bibfnamefont {I.}~\bibnamefont {Bloch}}, \ and\ \bibinfo {author}
  {\bibfnamefont {S.}~\bibnamefont {Kuhr}},\ }\href {\doibase
  10.1038/nature09378} {\bibfield  {journal} {\bibinfo  {journal} {Nature}\
  }\textbf {\bibinfo {volume} {467}},\ \bibinfo {pages} {68} (\bibinfo {year}
  {2010})}\BibitemShut {NoStop}%
\bibitem [{\citenamefont {Mikeska}\ and\ \citenamefont
  {Kolezhuk}(2004)}]{Mikeska2004}%
  \BibitemOpen
  \bibfield  {author} {\bibinfo {author} {\bibfnamefont {H.}~\bibnamefont
  {Mikeska}}\ and\ \bibinfo {author} {\bibfnamefont {A.}~\bibnamefont
  {Kolezhuk}},\ }\enquote {\bibinfo {title} {One-dimensional magnetism},}\ \
  (\bibinfo {year} {2004})\ pp.\ \bibinfo {pages} {1--83}\BibitemShut {NoStop}%
\bibitem [{\citenamefont {Luttinger}(1963)}]{Luttinger1963}%
  \BibitemOpen
  \bibfield  {author} {\bibinfo {author} {\bibfnamefont {J.~M.}\ \bibnamefont
  {Luttinger}},\ }\href {\doibase 10.1063/1.1704046} {\bibfield  {journal}
  {\bibinfo  {journal} {Journal of Mathematical Physics}\ }\textbf {\bibinfo
  {volume} {4}},\ \bibinfo {pages} {1154} (\bibinfo {year} {1963})},\ \Eprint
  {http://arxiv.org/abs/https://doi.org/10.1063/1.1704046}
  {https://doi.org/10.1063/1.1704046} \BibitemShut {NoStop}%
\bibitem [{\citenamefont {Mattis}\ and\ \citenamefont {Lieb}(1965)}]{Lieb1965}%
  \BibitemOpen
  \bibfield  {author} {\bibinfo {author} {\bibfnamefont {D.~C.}\ \bibnamefont
  {Mattis}}\ and\ \bibinfo {author} {\bibfnamefont {E.~H.}\ \bibnamefont
  {Lieb}},\ }\href {\doibase 10.1063/1.1704281} {\bibfield  {journal} {\bibinfo
   {journal} {Journal of Mathematical Physics}\ }\textbf {\bibinfo {volume}
  {6}},\ \bibinfo {pages} {304} (\bibinfo {year} {1965})},\ \Eprint
  {http://arxiv.org/abs/https://doi.org/10.1063/1.1704281}
  {https://doi.org/10.1063/1.1704281} \BibitemShut {NoStop}%
\bibitem [{\citenamefont {Giamarchi}(2003)}]{Giamarchi2003}%
  \BibitemOpen
  \bibfield  {author} {\bibinfo {author} {\bibfnamefont {T.}~\bibnamefont
  {Giamarchi}},\ }\href
  {https://www.oxfordscholarship.com/view/10.1093/acprof:oso/9780198525004.001.0001/acprof-9780198525004}
  {\emph {\bibinfo {title} {Quantum Physics in One Dimension.}}}\ (\bibinfo
  {publisher} {Oxford University Press},\ \bibinfo {year} {2003})\ pp.\
  \bibinfo {pages} {160--199}\BibitemShut {NoStop}%
\bibitem [{\citenamefont {Hallberg}\ \emph {et~al.}(1995)\citenamefont
  {Hallberg}, \citenamefont {Horsch},\ and\ \citenamefont
  {Mart\'{\i}nez}}]{Hallberg1995}%
  \BibitemOpen
  \bibfield  {author} {\bibinfo {author} {\bibfnamefont {K.~A.}\ \bibnamefont
  {Hallberg}}, \bibinfo {author} {\bibfnamefont {P.}~\bibnamefont {Horsch}}, \
  and\ \bibinfo {author} {\bibfnamefont {G.}~\bibnamefont {Mart\'{\i}nez}},\
  }\href {\doibase 10.1103/PhysRevB.52.R719} {\bibfield  {journal} {\bibinfo
  {journal} {Phys. Rev. B}\ }\textbf {\bibinfo {volume} {52}},\ \bibinfo
  {pages} {R719} (\bibinfo {year} {1995})}\BibitemShut {NoStop}%
\bibitem [{\citenamefont {Goodfellow}\ \emph {et~al.}(2016)\citenamefont
  {Goodfellow}, \citenamefont {Bengio},\ and\ \citenamefont
  {Courville}}]{Goodfellow-et-al-2016}%
  \BibitemOpen
  \bibfield  {author} {\bibinfo {author} {\bibfnamefont {I.}~\bibnamefont
  {Goodfellow}}, \bibinfo {author} {\bibfnamefont {Y.}~\bibnamefont {Bengio}},
  \ and\ \bibinfo {author} {\bibfnamefont {A.}~\bibnamefont {Courville}},\
  }\href {http://www.deeplearningbook.org} {\emph {\bibinfo {title} {Deep
  Learning}}}\ (\bibinfo  {publisher} {MIT Press},\ \bibinfo {year}
  {2016})\BibitemShut {NoStop}%
\bibitem [{Sup()}]{Supplementary}%
\BibitemOpen
 See Supplemental Material below for information about the reconstruction of ground-state energies and densities from ML wave functions, performance of the DTWF for all $U/t$ values tested across the considered phase transition,  DTCF performance on the validation set and a detailed description of the finite-size scaling properties of the two-point density correlation functions and logarithmic corrections in the critical point, which includes Refs~\cite{Kaplan1987}
\BibitemShut {NoStop}
\bibitem [{ITe()}]{ITensor}%
  \BibitemOpen
  \href@noop {} {\bibinfo  {journal} {\mbox{ITensor Library} (version 2.0.11)
  http://itensor.org}\ }\BibitemShut {NoStop}%
\bibitem [{\citenamefont {Kaplan}\ \emph {et~al.}(1987)\citenamefont {Kaplan},
  \citenamefont {Horsch},\ and\ \citenamefont {Borysowicz}}]{Kaplan1987}%
  \BibitemOpen
\bibfield  {journal} {  }\bibfield  {author} {\bibinfo {author} {\bibfnamefont
  {T.~A.}\ \bibnamefont {Kaplan}}, \bibinfo {author} {\bibfnamefont
  {P.}~\bibnamefont {Horsch}}, \ and\ \bibinfo {author} {\bibfnamefont
  {J.}~\bibnamefont {Borysowicz}},\ }\href {\doibase 10.1103/PhysRevB.35.1877}
  {\bibfield  {journal} {\bibinfo  {journal} {Phys. Rev. B}\ }\textbf {\bibinfo
  {volume} {35}},\ \bibinfo {pages} {1877} (\bibinfo {year}
  {1987})}\BibitemShut {NoStop}%
\bibitem [{\citenamefont {Carleo}\ and\ \citenamefont
  {Troyer}(2017)}]{Carleo2017}%
  \BibitemOpen
  \bibfield  {author} {\bibinfo {author} {\bibfnamefont {G.}~\bibnamefont
  {Carleo}}\ and\ \bibinfo {author} {\bibfnamefont {M.}~\bibnamefont
  {Troyer}},\ }\href {\doibase 10.1126/science.aag2302} {\bibfield  {journal}
  {\bibinfo  {journal} {Science}\ }\textbf {\bibinfo {volume} {355}},\ \bibinfo
  {pages} {602} (\bibinfo {year} {2017})},\ \Eprint
  {http://arxiv.org/abs/https://science.sciencemag.org/content/355/6325/602.full.pdf}
  {https://science.sciencemag.org/content/355/6325/602.full.pdf} \BibitemShut
  {NoStop}%
\end{thebibliography}
\end{document}


\title{Supplementary materials}
\author{Javier Robledo Moreno}
\email{jrm874@nyu.edu}
\affiliation{
Center for Computational Quantum Physics, Flatiron Institute, New York, NY 10010 USA
}
\affiliation{Center for Quantum Phenomena, Department of Physics, New York University, 726 Broadway, New York, New York 10003, USA
}

\author{Giuseppe Carleo}
\email{gcarleo@flatironinstitute.org}
\affiliation{
Center for Computational Quantum Physics, Flatiron Institute, New York, NY 10010 USA
}

\author{Antoine Georges}
\email{ageorges@flatironinstitute.org}
\affiliation{
Center for Computational Quantum Physics, Flatiron Institute, New York, NY 10010 USA
}
\affiliation{Coll{\`e}ge de France, 11 place Marcelin Berthelot, 75005 Paris, France}
\affiliation{CPHT, CNRS, {\'E}cole Polytechnique, IP Paris, F-91128 Palaiseau, France}
\affiliation{DQMP, Universit{\'e} de Gen{\`e}ve, 24 quai Ernest Ansermet, CH-1211 Gen{\`e}ve, Suisse}

\date{\today}

\maketitle

\section{Description of the non linearity that forces the normalization of the predicted wave function}
As described in the main text, the network to predict the DTWF map takes as an input the $N$ values of the density on each lattice site and outputs the ground-state wave function components in the occupation basis. All the layers are connected by the composition of an affine transformation and a nonlinear rectifier function, $\textrm{Relu}(x)=\mathrm{max}(0,x)$, except for the output layer. We chose the non linearity of this layer to force the normalization of the wave function. Specifically, if $\vec{h}^{(D)}$ is a vector containing the units of the last hidden layer and $\vec{\psi}$ a vector containing the wave function components, 
    we have:
    \begin{equation}\label{Eq_05: Normalized layer of the network}
        \vec{\psi} = \frac{  W^{(D)} \cdot \vec{h}^{(D)} + \vec{C}^{(D)}}{\sqrt{ |W^{(D)} \cdot \vec{h}^{(D)} + \vec{C}^{(D)}|^2}},
    \end{equation}
    where $W^{(D)}$ and $\vec{C}^{(D)}$ are the weights and biases of the affine transformation corresponding to the last layer of the network. This way, the predicted wave function is automatically normalized.

\section{Reconstructing ground state energies and densities from ML constructed wave functions.}
In this section we show that the ML constructed wave functions accurately predict the ground-state energy as well as the density distributions they are constructed from. In order to show it, for the largest system size analyzed ($N = 14$), we take all the examples from the validation set and construct their wave functions from their respective density distributions. The predicted wave functions are used to compute the density distribution and energy of the system. The distance between the target ($\rho^{\textrm{Tar}}$) and predicted ($\rho^{\textrm{Pred}}$) density distributions is defined as:
\begin{equation}\label{EQ 1: Density distance}
    D \left( \rho^{\textrm{Tar}}, \rho^{\textrm{Pred}} \right) = \sum_{i = 1}^N \left|\rho_i^{\textrm{Tar}}-\rho_i^{\textrm{Pred}} \right| ^2.
\end{equation}

Similarly, the relative error in the predicted energy of the ground state ($E_0^{\textrm{Pred}}$) with respect to the true value ($E_0^{\textrm{Tar}}$) is defined as:
\begin{equation}\label{EQ 2: Energy distance}
    \epsilon \left(E_0^{\textrm{Tar}}, E_0^{\textrm{Pred}} \right) = \frac{\left|E_0^{\textrm{Tar}}- E_0^{\textrm{Pred}}\right|}{\left| E_0^{\textrm{Tar}} \right|}.
\end{equation}

    \begin{figure*}[t]
        \includegraphics[width=1\textwidth]{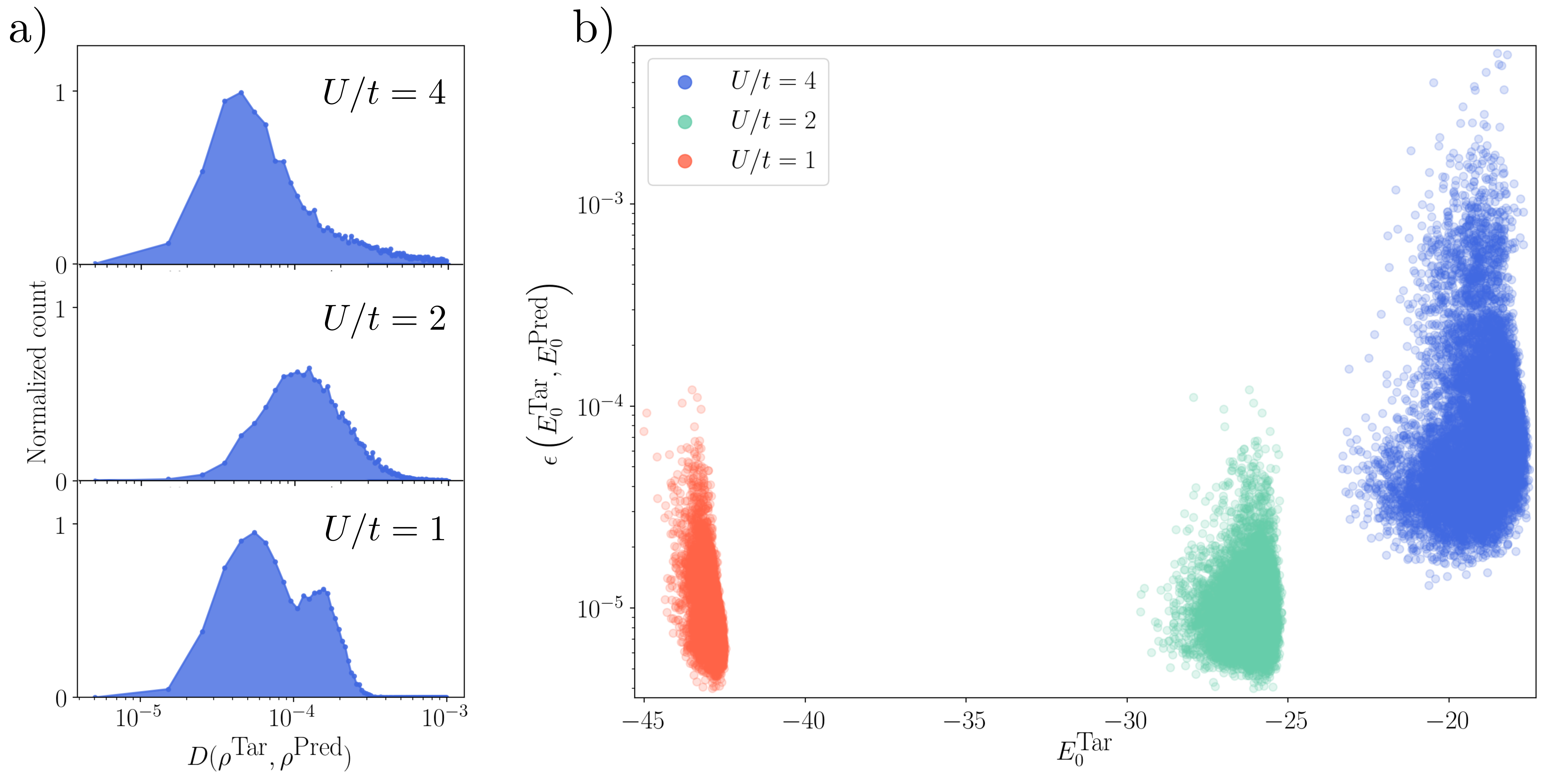}
        \caption{\label{FIG_S1: Sanity test} Self-consistency tests in the larger lattice analyzed (N = 14). \textbf{a)} Histograms of the distance between target and predicted densities in all of the examples of the validation set as defined in Eq.~(\ref{EQ 1: Density distance}). Data corresponding to $U/t = 1$, $U/t = 2$ and $U/t = 4$ is displayed. \textbf{b)} Scatter plot of the error in the predicted ground-state energy (Eq.~(\ref{EQ 2: Energy distance})) as a function of the target energy for $U/t = 1$, $U/t = 2$ and $U/t = 4$. Each point corresponds to one sample in the validation set.} 
    \end{figure*} 

Figure~\ref{FIG_S1: Sanity test} a) shows histograms of the distance between target and predicted densities over the validation set for $U/t = 1$, $U/t = 2$ and $U/t = 4$. The distance between density distributions is peaked around $D(\rho^{\textrm{Tar}}, \rho^{\textrm{Pred}}) \sim 10^{-4}$, showing that the predicted wave functions can self-consistently reconstruct the density they are constructed from. Figure~\ref{FIG_S1: Sanity test} b) shows the relative error of the energy predicted from the ML constructed wave function as a function of the target energy for different $U/t$ values. The metallic phase and critical point show typical errors no larger than $\epsilon(E_0^{\textrm{Tar}}, E_0^{\textrm{Pred}}) \sim 10^{-4}$ and the Mott phase no larger than $\epsilon(E_0^{\textrm{Tar}}, E_0^{\textrm{Pred}}) \sim 5\cdot10^{-3}$. This illustrates that the constructed wave functions can self-consistently describe observables of the system, as expected from their high fidelity (see Figure~1 in the main text).

\section{Performance in structured potentials for different $U/t$ values.}
\begin{figure*}[t]
        \includegraphics[width=1\textwidth]{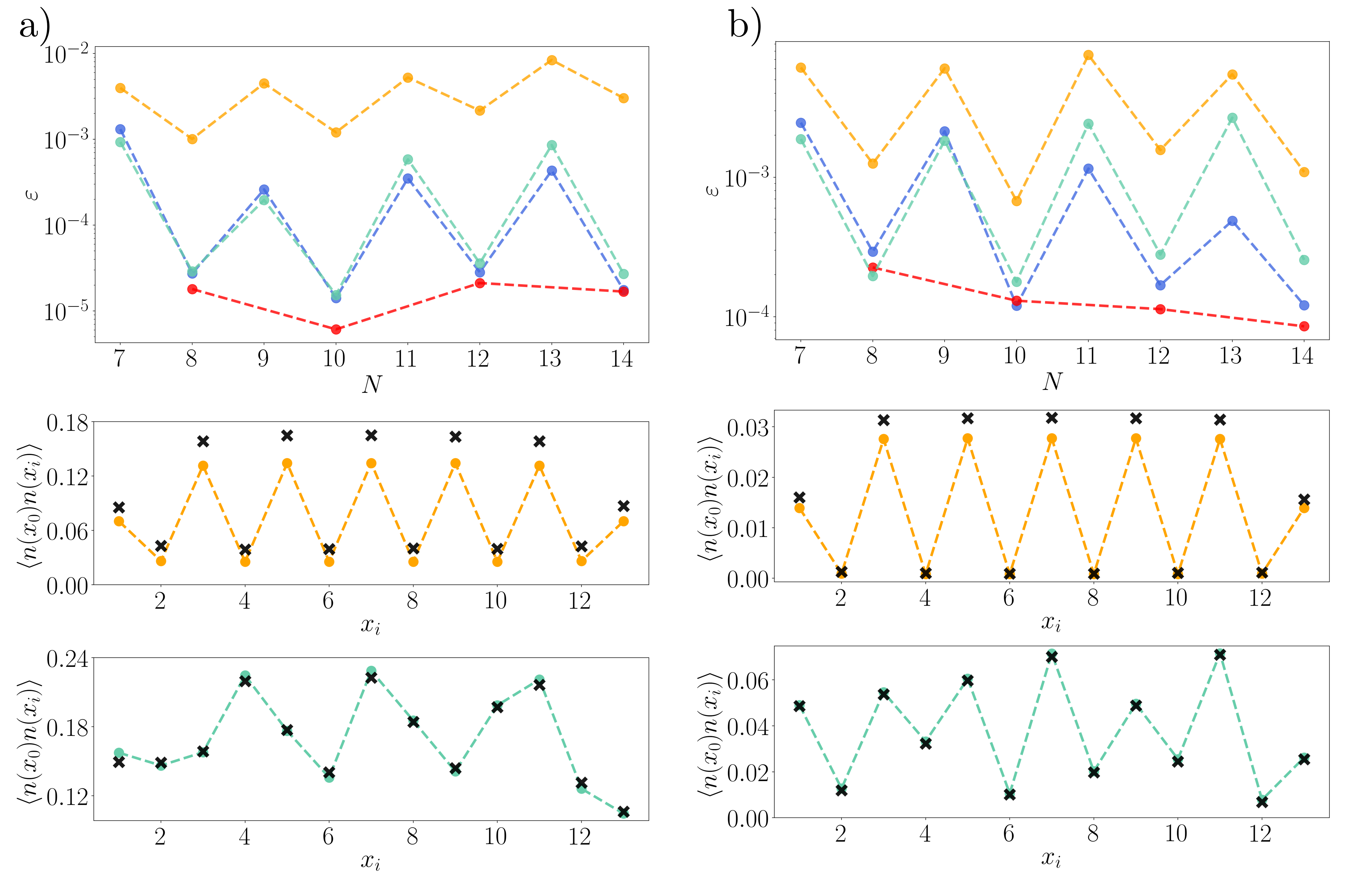}
        \caption{\label{FIG_S2: Accuracy_ext_pot} Performance of the network in structured potentials. As in the main text, the tested potentials are quadratic (blue), no potential (red), periodic with period $N/4$ (green) and staggered (orange). \textbf{a)} \textbf{U/t = 1} Top: Infidelity of the predicted wave function as a function of the system size. Middle: Two-point density correlation functions computed from exact (dots connect by dashed lines) and ML predicted (black crosses) wave functions in the staggered potential case. Bottom: Same as the middle panel but for the periodic with period $N/4$ potential. \textbf{b)} \textbf{U/t = 4} Same as in panel a).} 
    \end{figure*}

The aim is this section is to test the performance of the network in structured potentials for different values of $U/t$ not shown in the main text. Figure~\ref{FIG_S2: Accuracy_ext_pot} a) and b) display the infidelity of the predicted wave function as a function of system size for $U/t = 1$ and $U/t = 4$ for the same collection of structured potentials shown in the main text (see Figure~2). The infidelity in the staggered potential case is up to two orders of magnitude larger than its value for the rest of the potentials in the metallic phase and one order of magnitude larger in the Mott phase. These results show that the performance of the network suffers in the presence of a quantum phase transition (QPT) regardless of the $U/t$ value. 

Bottom panels of Figure~\ref{FIG_S2: Accuracy_ext_pot} a) and b) display two-point density correlation functions computed from exact and predicted wave functions in the staggered and the periodic with period $N/4$ potential cases. It shows that the larger infidelity of the staggered potential wave function translates to significant errors when computing observables.

\section{Testing DTCF map on the validation set for $N = 18$ lattice sites.}
In this section we test the performance of the network that approximates the density to correlation function (DTCF) map over the validation set, consisting on $10^4$ examples, for each value of $U/t$ analyzed in the main text. System size $N = 18$ is analyzed. The neural network correlators $\left(\langle n(x_i)n(x_j) \rangle_{\textrm{Pred}}\right)$ are compared to the target values $\left(\langle n(x_i)n(x_j) \rangle_{\textrm{Tar}}\right)$ by computing the absolute error for each example in the validation set:
\begin{equation}\label{EQ 3: Correlations error}
    \epsilon \left(\langle n(x_i)n(x_j) \rangle \right) = \left|\langle n(x_i)n(x_j) \rangle_{\textrm{Pred}} - \langle n(x_i)n(x_j) \rangle_{\textrm{Tar}} \right|,
\end{equation}
for every $x_i$ and $x_j$ pair. Within the same example, the errors corresponding to pairs separated by the same distance in the lattice ($|x_i-x_j|$) are averaged together. Provided that the lattice has periodic boundary conditions, the maximum distance between lattice sites is $|x_i-x_j| = a\cdot N/2$, where $a$ is the lattice constant.

\begin{figure*}[t]
        \includegraphics[width=1\textwidth]{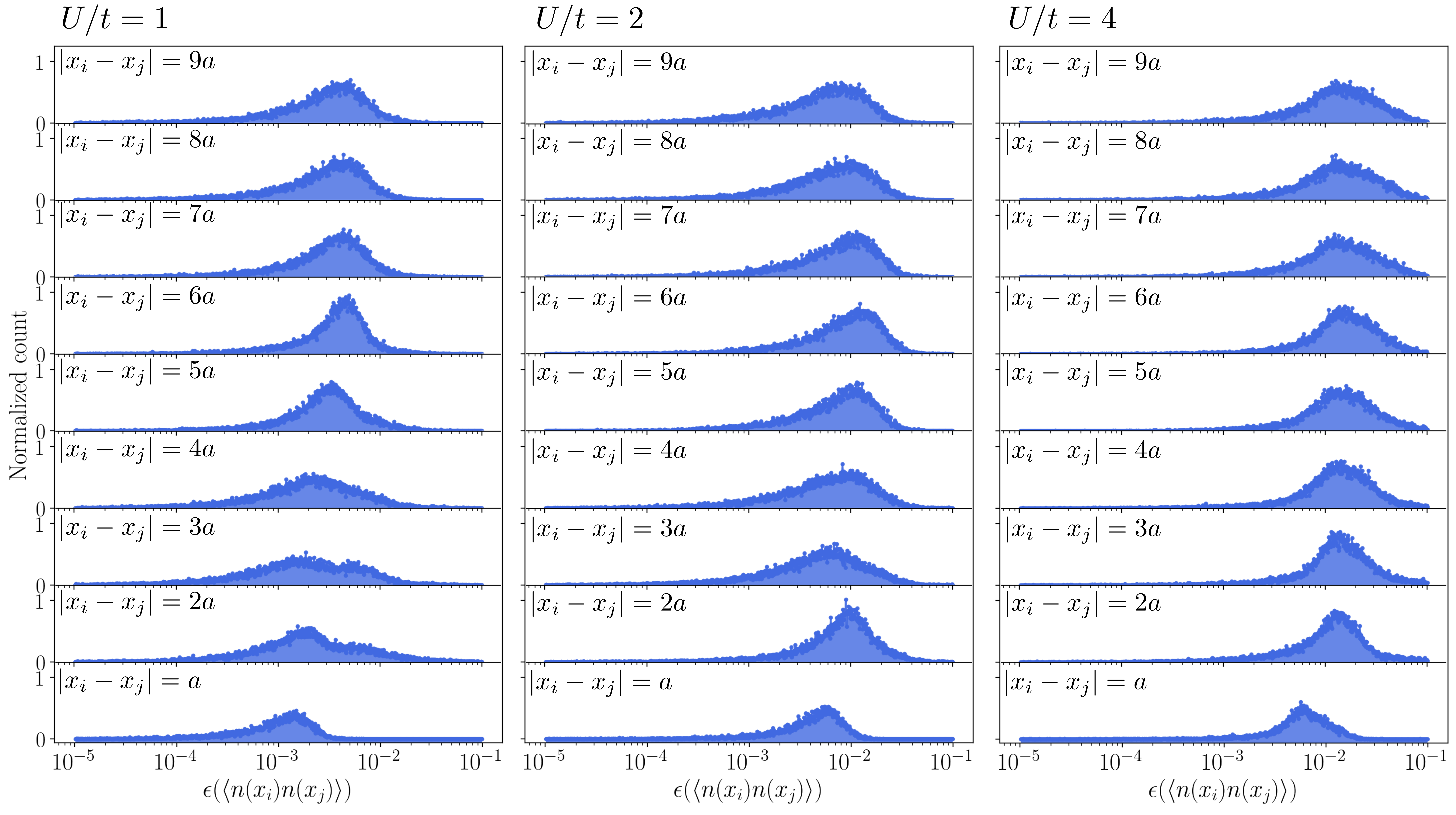}
        \caption{\label{FIG_S3: Conv_net_histograms} Normalized histograms of error over the validation set, as defined in Eq.~(\ref{EQ 3: Correlations error}), when predicting density-density correlators for the different $U/t$ values analyzed. The system has $N=18$ lattice sites. For each $U/t$ value, different histograms display the errors for lattice site pairs separated by distance $|x_i-x_j|$.} 
    \end{figure*}

Figure~\ref{FIG_S3: Conv_net_histograms} shows histograms of the error when constructing the DTCF map at different values of $U/t$. As mentioned above, the errors are grouped by the distance between $x_i$ and $x_j$. Histograms for each distance are shown for each $U/t$ value. Error does not increase significantly with the distance between lattice site pairs. The error increases, but not significantly, as the correlation increases with $U/t$.

\section{Testing DTCF map on the validation set for $N = 50$ lattice sites.}  

    In order to argue that the DTWF map construction we propose does not deteriorate with system size we tested its performance in a significantly larger system, with $N = 50$ system sites. We studied the construction of the map in systems with $U/t = 1$, $U/t = 2$ and $U/t = 4$, as in the main text. The training sets were generated following the same procedure described in the main text and the ground-state is found using the Density Matrix Renormalization Group technique on ITensor. Once the ground-state is constructed, correlations are measured using standard tensor-network contractions. 
    
\begin{figure*}[t]
        \includegraphics[width=1\textwidth]{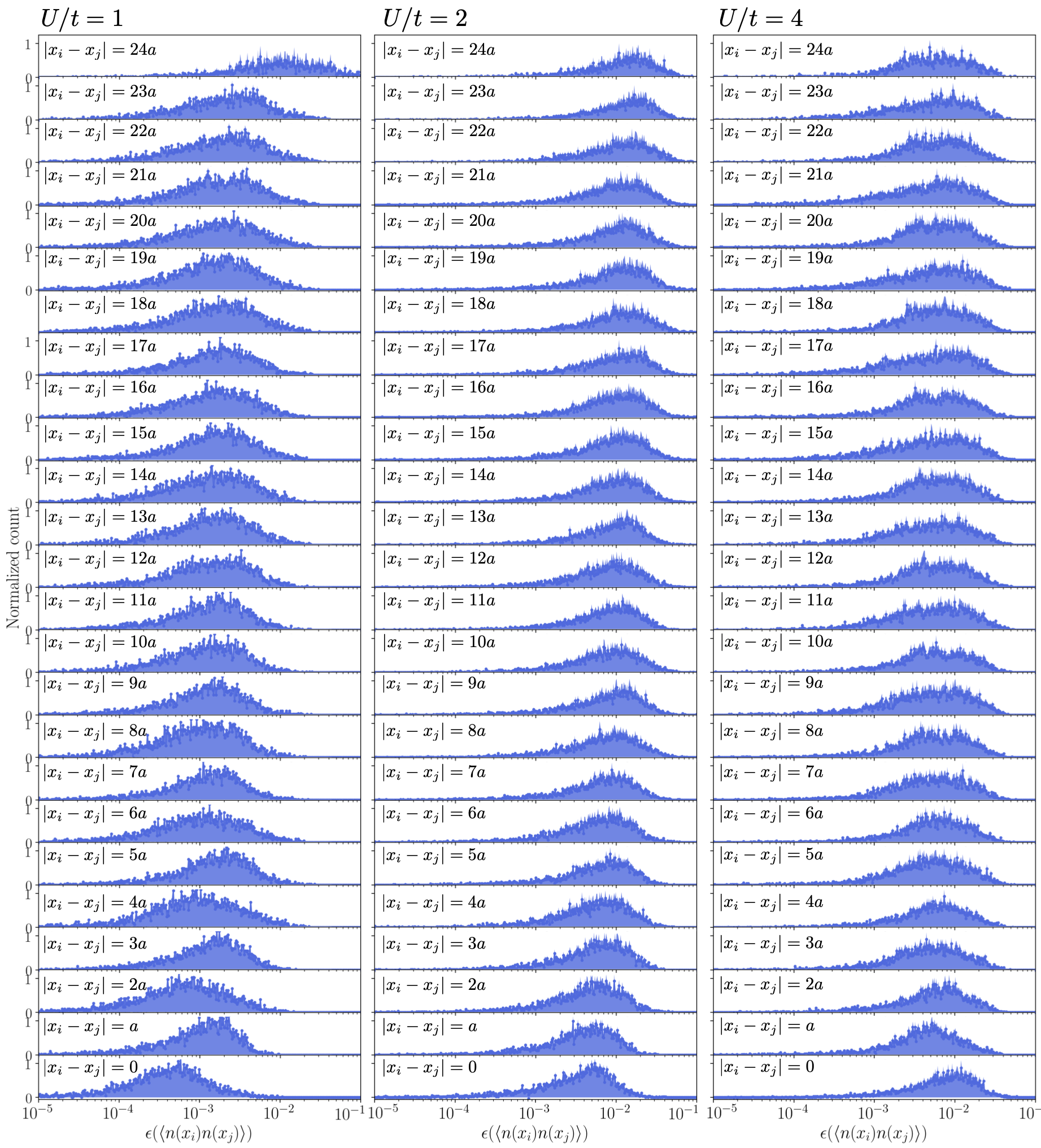}
        \caption{\label{FIG_S4: Conv_net_histogramsDMRG} Normalized histograms of error over the validation set, as defined in Eq.~(\ref{EQ 3: Correlations error}), when predicting density-density correlators for the different $U/t$ values analyzed. The system has $N=50$ lattice sites. For each $U/t$ value, different histograms display the errors for lattice site pairs separated by distance $|x_i-x_j|$.} 
    \end{figure*}

    The training set contains 80,000 examples for each $U/t$ value and the performance is tested in a validation set with 20,000 examples. Figure~\ref{FIG_S4: Conv_net_histogramsDMRG} shows histograms of the error when constructing the DTCF map at different values of $U/t$. As in the previous section, the errors are grouped by the distance between $x_i$ and $x_j$ and histograms for each distance are shown for each $U/t$ value. The error does not increase significantly with the distance between site pairs. Most importantly, these error distributions are similar to the ones in the system with $N = 18$ lattice sites (see Figure~\ref{FIG_S4: Conv_net_histogramsDMRG}), showing that the proposed method to infer correlation functions from the local density does not show a performance decrease as the number of lattice sites in increased.

\section{Luttinger Liquid parameter, finite-size scaling of correlation functions and logarithmic corrections.}

    In this section we perform a similar analysis to it in~\cite{Hallberg1995} to obtain the finite-size scaling of the connected correlation functions in the metallic phase and critical point. The finite-size scaling analysis gives us access to the logarithmic correction in the critical point.  We will  study the finite-size effects in the correlation functions using DMRG to find the scaling functions and to show that logarithmic corrections can be in principle captured with this approach. Then, we will repeat the same analysis with the machine-learning correlations. 
    
    \subsection{Definitions and expected scaling}
    
    In the absence of an external potential the correlation functions of this system are exactly known in the thermodynamic limit from Luttinger Liquid (LL) theory~\cite{Giamarchi2003}. In the metallic phase they take the form:
    \begin{equation}\label{Eq: correlations LL}
        C(l) = \langle n(x_i) n(x_{i+l}) \rangle -\frac{1}{4} = C_1\frac{1}{l^2} + C_2(-1)^l \left(\frac{1}{l} \right)^{2K}
    \end{equation}
     where $K$ is the Luttinger Liquid parameter exactly known from Bethe ansatz:
    \begin{equation}\label{Eq: LL parameter}
        \frac{1}{K} = \frac{2}{\pi}\arccos\left( -\frac{J_z}{J_{xy}}\right).
    \end{equation}
    In the critical point the staggered term acquires a logarithmic correction:
    \begin{equation}\label{Eq: correlations critical point}
        C(l) = C_1\frac{1}{l^2} + C_2(-1)^l \frac{\sqrt{\log (cl)}}{l}.
    \end{equation}
    where the constant $c$ is estimated to be $c = 23.21$~\cite{Hallberg1995}. This constant is quite large for the system sizes under consideration. This means that the logarithmic corrections are very subtle, so chains with a few thousand lattice sites would be necessary to capture this logarithmic corrections.
    
    In the Mott phase the correlation functions are known to have an exponential decay superimposed to the power-law decay but there is no known close form~\cite{Giamarchi2003}.
    
\begin{figure*}[t]
        \includegraphics[width=.8\textwidth]{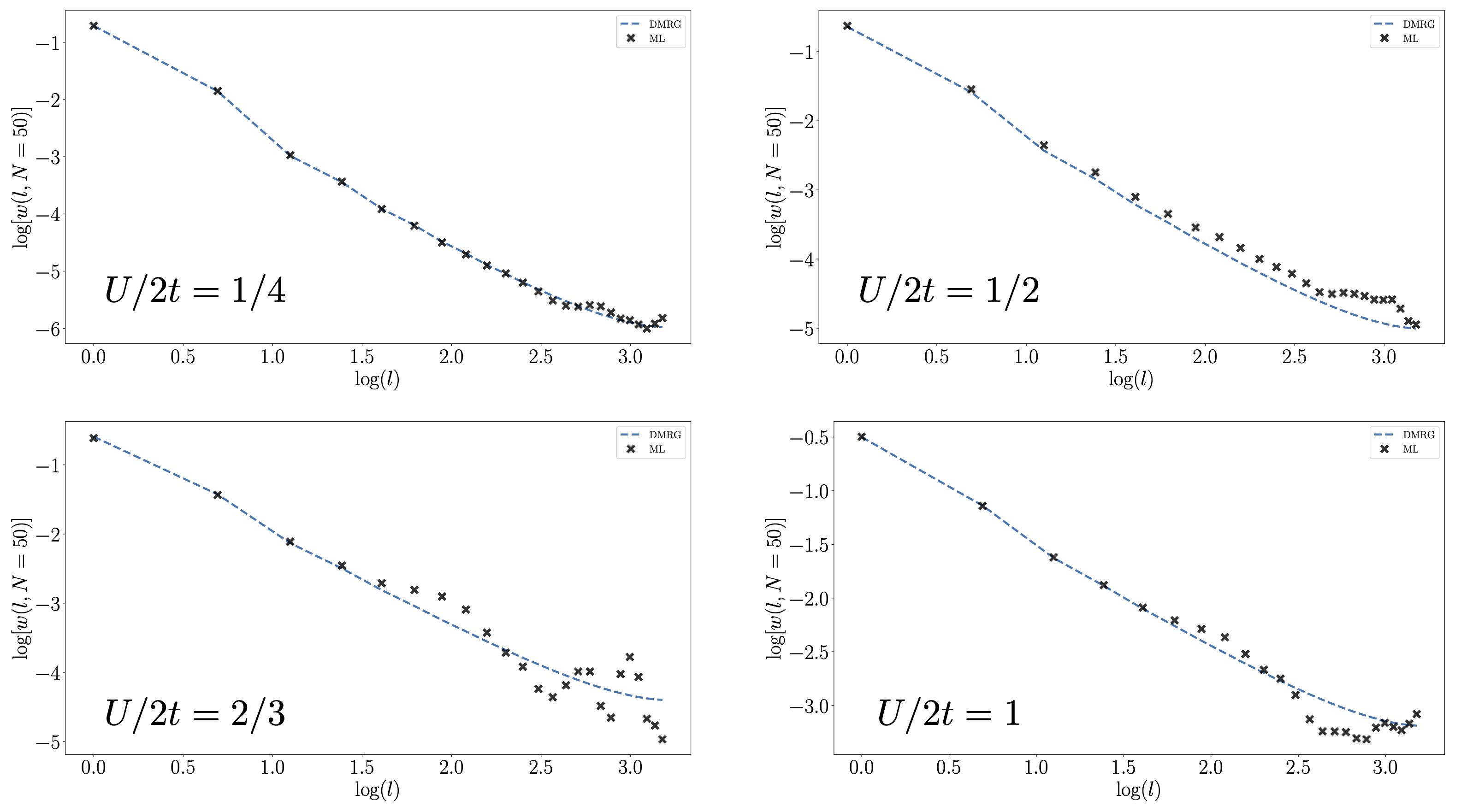}
        \caption{\label{FIG_S5: Raw correlations for LL parameter} Raw average correlation functions (Eq~\ref{Eq: average correlator}) as a function of the distance between correlation pairs. Different panels show the data at different values of the interaction in the metallic phase and critical point. Black crosses are the machine learning predicted correlation functions and the discontinuous blue line are the correlations from DMRG.} 
    \end{figure*}

    We are interested in the staggered term of the correlation functions proportional to $C_2$, therefore it is better to consider the staggered correlation functions: $w(l,N) = (-1)^lC(l)$. In particular, in order to remove the $1/l^2$ background, we will consider the average correlation functions:
    \begin{equation}\label{Eq: average correlator}
        \bar{w}(l,N) = \frac{1}{4} \left[w(l-1,N) + 2 w(l,N) + w(l+1,N) \right],
    \end{equation}
    where the $1/l^2$ term is now of order $\mathcal{O}(1/l^4)$, making it negligible (for sufficiently large $l$, $l = 5$ is enough) compared to the term proportional to $1/l^{2K}$ that we are interested in. With this redefinition of the correlation functions, the average correlations take the form:
    \begin{equation}\label{Eq: average correlations power-law}
        \bar{w}(l,N) \sim \frac{1}{l^{2K}}; \;\;\;\;\bar{w}(l,N) \sim \frac{\sqrt{\log (cl)}}{l} ,
    \end{equation}
    
    in the metallic phase and critical point respectively. We extract the LL parameter directly from these expressions by fitting the $\log(\bar{w}(l,N))$ vs $\log(l)$ to a straight line. This is the method we use to obtain the LL parameter shown in Figure 3 c) in the main text, using the largest system size available to us $N = 50$. The raw data  of $\log(\bar{w}(l,N = 50))$ vs $\log(l)$ is shown in Figure~\ref{FIG_S5: Raw correlations for LL parameter}. It shows that the neural network is predicting very accurate correlation functions that can capture the correct power-law decay. Even though the decay of the correlations is consistent with a power-law decay at this system size, there are deviations from pure power-law for large values of $l$ in all interaction strengths. This is due to finite-size effects. The logarithmic corrections cannot be captured in this system size as discussed above. In order to obtain the pure power-law decay of the correlations in the metallic phase and the logarithmic corrections we do a finite-size scaling analysis.
    
    Thanks to the expected pure power-law behaviour (almost pure in the critical point) we use self similarity arguments to derive the expected finite-size scaling of the correlation functions. If we  transform coordinates such that $l\rightarrow  l/T$ and therefore $N\rightarrow N/T$ and the system has dilation symmetry, then, it is expected that $\bar{w}(l,N)\rightarrow \bar{w}(l,N) T^{\Delta w}$. Therefore, we have that $\bar{w}(l,N)  = T^{-\Delta w} \bar{w}(l/T,N/T)$ must be satisfied upon dilation. Choosing $T$ such that $l/T =1$, it must be satisfied that:
    \begin{equation}\label{Eq: dilation sym scaling function}
        \bar{w}(l,N) = l^{-\Delta w} \phi (l/N),
    \end{equation}
    where $\phi$ is the scaling function in this case. From~\cite{Hallberg1995} we know that the scaling function is the scaling function of the $XY$ model raised to power $\alpha$:
    \begin{equation}\label{Eq: scaling relation}
        \bar{w}(l,N) = \bar{w}(l,\infty) (f_\textrm{XY}(l/N))^\alpha,
    \end{equation}
    where $\bar{w}(l,\infty)$ are the correlation functions in the thermodynamic limit. $f_\textrm{XY}(l/N) =  1+0.28822 \sinh^2 (1.673 \cdot l/N)$ from a phenomenological fit from~\cite{Kaplan1987}. In conclusion, the proposed scaling should be satisfied in the metallic phase and critical point (because logarithmic corrections are small), and should fail in the Mott phase.
    
    The idea is to use this scaling behaviour to obtain the extrapolated correlation functions in the thermodynamic limit $\bar{w}(l,\infty)$ and obtain the LL parameter in the metallic phase and the logarithmic corrections in the critical point. 
    
    \subsection{DMRG analysis}
\begin{figure*}[t]
        \includegraphics[width=1\textwidth]{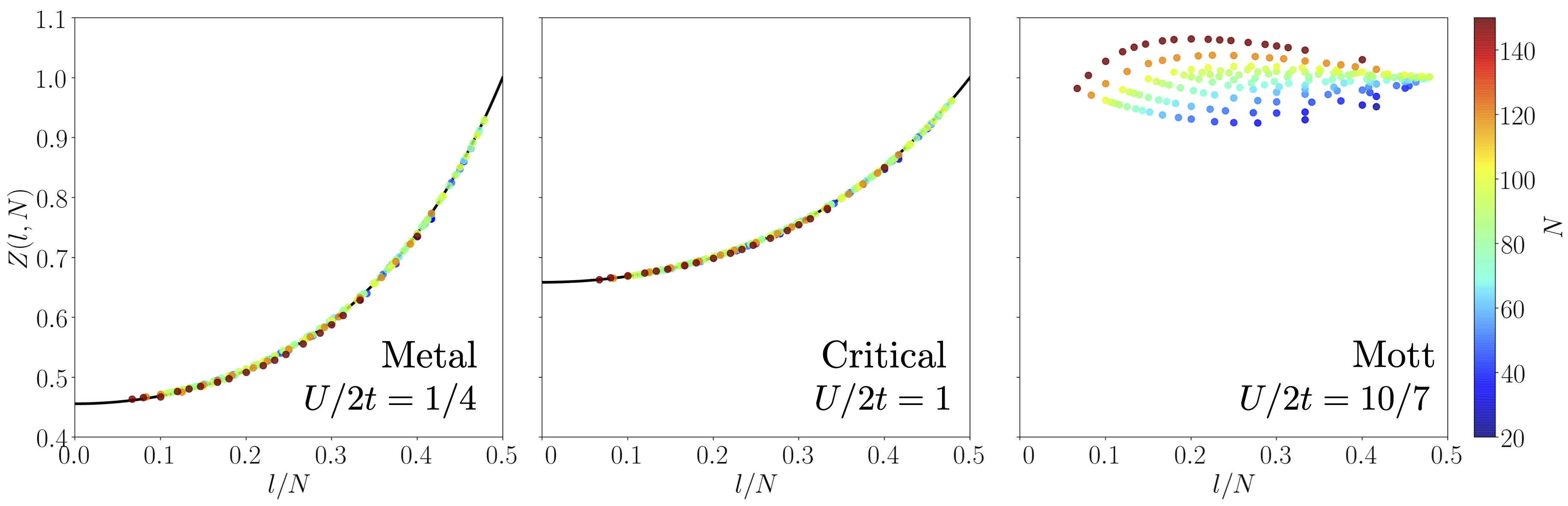}
        \caption{\label{FIG_S6: DMRG data collapse to scaling function} Ratio $Z(l,N)$ as defined in Eq.~\ref{Eq: Z ratio} for different values of the interaction corresponding to different phases as indicated in each panel. The data points come from DMRG correlation functions. The solid black lines are the fit to scaling function~\ref{Eq: scaling relation} to find exponent $\alpha$. The color of the points indicates the system size they come from. The Mott case shows that data collapse is not possible because correlations are not pure power-law.} 
    \end{figure*}

    In order to find the scaling function and the exponent $\alpha$ we compute the correlations functions at different values of $U/2t$: $U/2t = 1/4$, $U/2t = 1$ and  $U/2t = 10/7$ (metallic, critical and Mott respectively), and $20$ different system sizes from $N = 20$ to $N = 150$. The ratio:
    \begin{equation}\label{Eq: Z ratio}
        Z(l,N) = \frac{\bar{w}(l,N)}{\bar{w}(l,2l)},
    \end{equation}
    should only depend on the scaling function and therefore be a function of $l/N$, only if the correlations are purely power law. Figure~\ref{FIG_S6: DMRG data collapse to scaling function} shows the ratio $Z(l,N)$ in the metallic phase, critical point and Mott phase as a function of $l/N$ from DMRG data. In the metallic phase and critical point the data collapse is almost perfect. The logarithmic correction in the critical point does not prevent data collapse because it is very small. The $Z(l,N)$ ratio does not collapse to a single smooth curve in the Mott phase as the correlations are not power-law. The data collapse can be used to extract the exponent $\alpha$ that determines the scaling function. We find $\alpha = 1.85$ in the critical point (consistent with $\alpha = 1.805$ found in~\cite{Hallberg1995}) and $\alpha = 3.48$ for $U/2t = 1/4$.
    
\begin{figure*}[t]
        \includegraphics[width=1\textwidth]{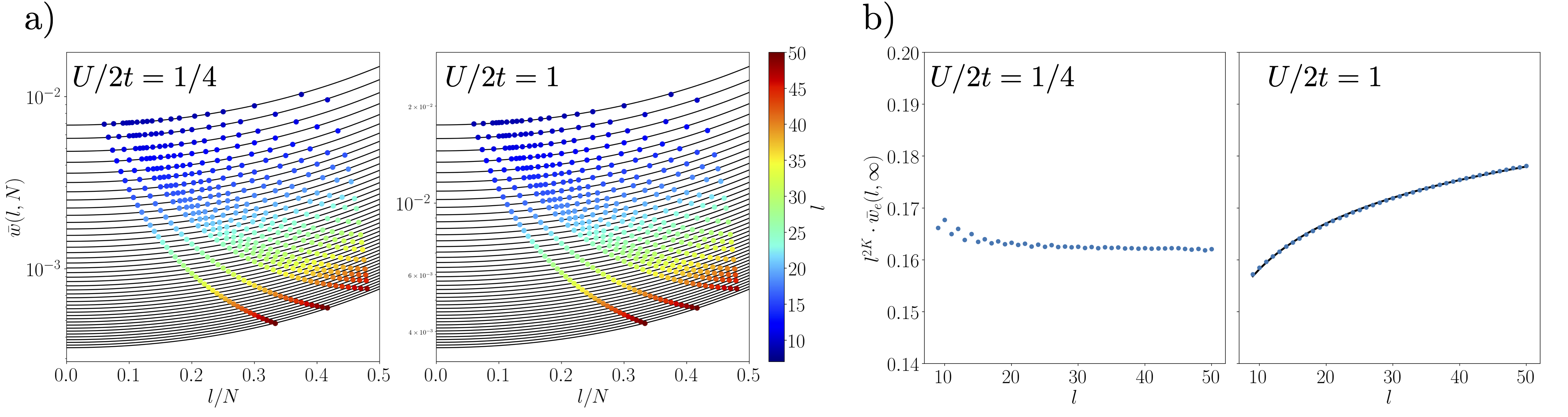}
        \caption{\label{FIG_S7: Extrapolated correlation functions and log correction} DMRG extrapolation of correlation functions in the thermodynamic limit \textbf{a)} Average correlations in the metallic phase and critical point as a function of $l/N$ grouped by the distance between correlation pairs $l$. The black lines are the fit to the scaling function in Eq.~\ref{Eq: scaling relation}. The only free parameter is $\bar{w}_e(l,\infty)$. \textbf{b)} Extrapolated correlations multiplied by $l^{2K}$ as a function of $l$ in the metallic phase and critical point. The black line in the critical point is the fit to $a \cdot \sqrt{\log(cl)}$ with $a$ and $c$ as fitting parameters.} 
    \end{figure*}

    Once the scaling function is known we extrapolate the correlation functions in the thermodynamic limit $\bar{w}_e(l,\infty)$ by fitting $w(l,N)$ as a function of $l/N$ to the scaling relation of Eq.~\ref{Eq: scaling relation} with $\bar{w}_e(l,\infty)$ as the only fitting parameter, for every value of $l$ separately. Figure~\ref{FIG_S7: Extrapolated correlation functions and log correction} a) shows this fit, demonstrating a good agreement with the proposed scaling function. We also show $l^{2K} \cdot \bar{w}_e(l,\infty)$ as a function of $l$ in the metallic phase and critical point in Figure~\ref{FIG_S7: Extrapolated correlation functions and log correction} b). The value of $K$ is taken from the Bethe ansatz solution of Eq.~\ref{Eq: LL parameter}. This quantity should be a constant in the metallic phase and increase as $a \cdot \sqrt{\log(cl)}$ in the critical point, as expected from Eq.~\ref{Eq: average correlations power-law}. In Figure~\ref{FIG_S7: Extrapolated correlation functions and log correction} b), the quantity $l^{2K} \cdot \bar{w}_e(l,\infty)$ approaches a constant value in the metallic phase and increases in the critical point. In the critical point we fit the resulting curve to $a \cdot \sqrt{\log(cl)}$ and extract the value of $c = 21.5$, consistent with the value found in~\cite{Hallberg1995}.
    
    The DMRG study proves that the scaling proposed in Eq.~\ref{Eq: scaling relation} applies only in the metallic phase and critical point because the correlation functions are purely power-law and power-law with a very small correction respectively. In the Mott phase the data collapse fails proving that the correlations functions are not consistent with a power-law decay. With that scaling relation we find the extrapolated correlation functions in the thermodynamic limit and find that they are purely power-law in the metallic phase and capture the logarithmic corrections in the critical point.
    
    \subsection{ML analysis}
    
\begin{figure*}[t]
        \includegraphics[width=1\textwidth]{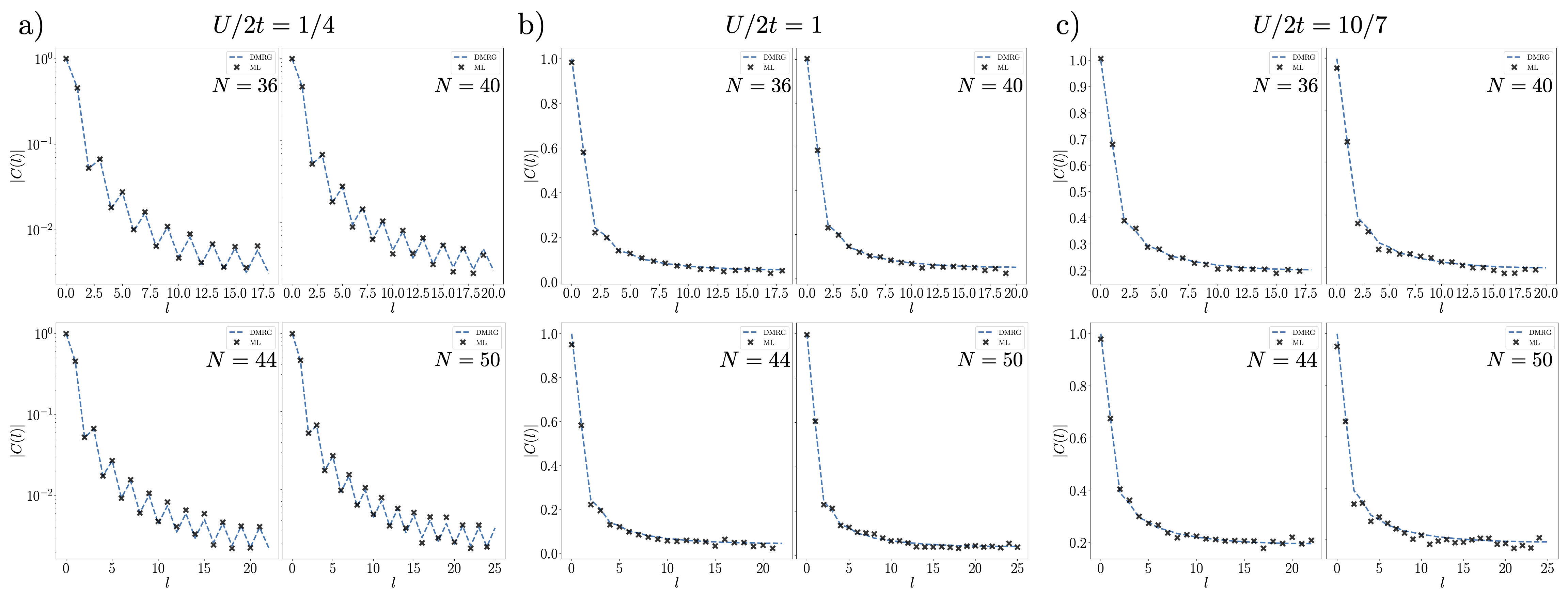}
        \caption{\label{FIG_S8:  ML raw data for finite-size scaling} Raw correlation-function data as a function of the distance between correlation pairs, used for the finite-size scaling analysis. Black crosses are the ML-predicted correlation functions and the dashed lines are the target values obtained from DMRG. Different panels correspond to different system sizes as indicated. \textbf{a)} $U/2t = 1/4$ metallic phase, \textbf{b)} $U/2t = 1$ critical point and \textbf{c)} $U/2t = 10/7$ Mott phase.} 
    \end{figure*}

    We do the same analysis with the ML-predicted correlations. In this case we have access to system sizes $N = 36,40,44,60$, and we will study the scaling behaviour of the correlation functions in the metallic phase, critical point and Mott phase ($U/2t = 1/4$, $U/2t = 1$ and $U/2t = 10/7$). In order to be able to carry out this analysis, very accurate ML predictions are required. Figure~\ref{FIG_S8:  ML raw data for finite-size scaling} shows the raw data used for the finite-size analysis. In all of the system sizes and values of the interaction, the prediction of each correlation pair is very accurate. This prediction does not seem to deteriorate as the interaction increases. 
    
\begin{figure*}[t]
        \includegraphics[width=1\textwidth]{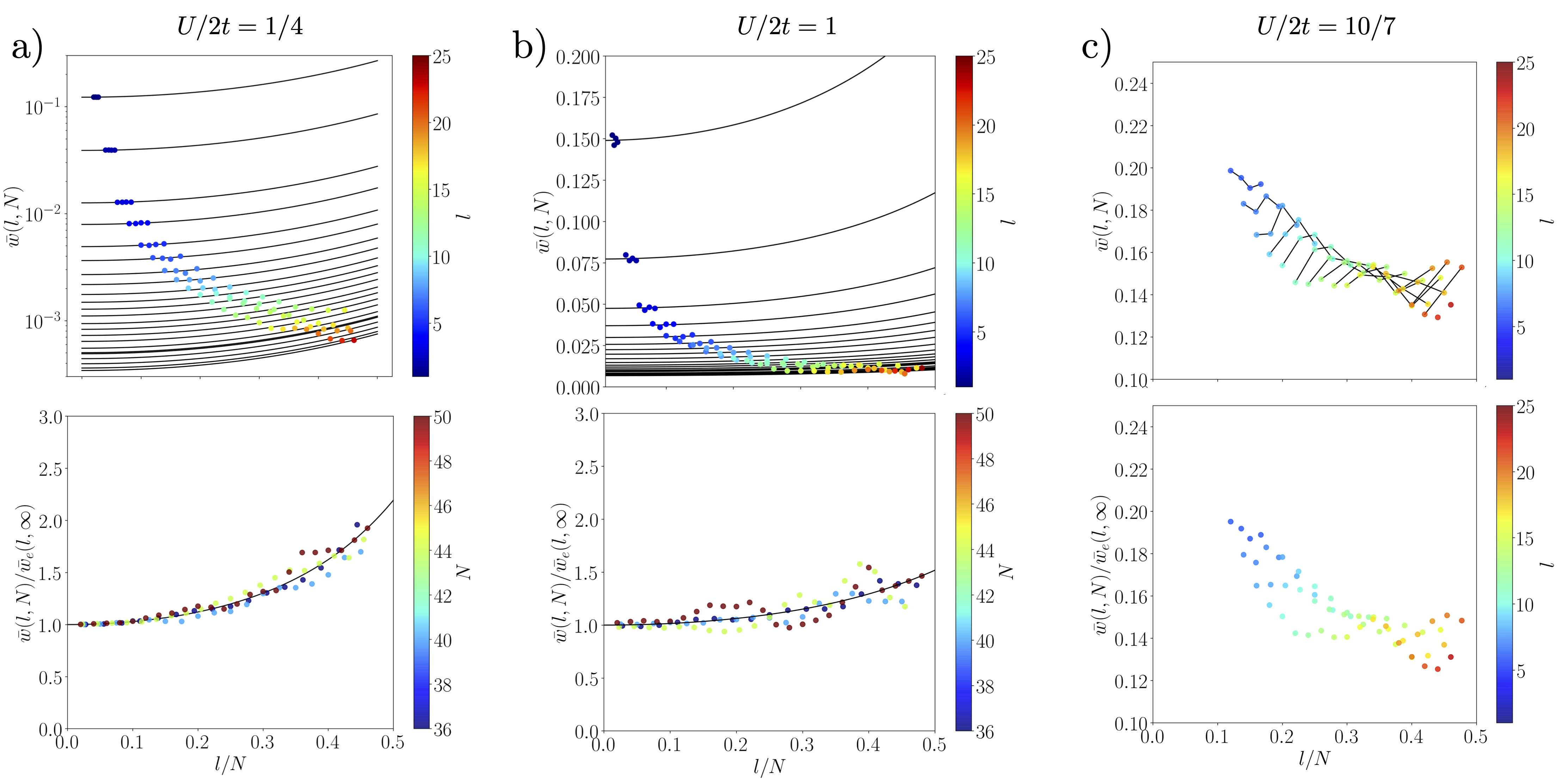}
        \caption{\label{FIG_S9:  ML extrapolation correlations and data collapse} Top panels show the average correlation functions as a function of $l/N$ for different values of $l$ as indicated by the color of the dots from ML correlations. The black curve in the metallic phase and critical point (a) and b)) is the fit to the known scaling function from the DMRG study (Figure~\ref{FIG_S6: DMRG data collapse to scaling function}). The black lines in c) are just for visual guidance. Bottom panels show the attempt to data collapse the correlation functions in the top panel. The solid black line in panels a) and b) are the known scaling functions. } 
    \end{figure*}

    Using the scaling function and values of $\alpha$ extracted from the DMRG study we extrapolate the value of the correlation functions in the thermodynamic limit. Top panels in Figure~\ref{FIG_S9:  ML extrapolation correlations and data collapse} a), b) and c) show the average value of correlation functions as a function of $l/N$ grouped by the distance between correlation pairs $l$. Panels a) and b) show the fit to the known scaling function to find the extrapolated value of the correlation functions in the thermodynamic limit. It shows that the fit is consistent with it found in the DMRG study. There are small deviations coming from small errors in the predicted correlations. With the extrapolated values we show (bottom panels of Figure~\ref{FIG_S9:  ML extrapolation correlations and data collapse}) that the quantity $\bar{w}(l,N)/\bar{w}_e(l,\infty)$ is consistent with a collapse to a smooth scaling function that only depends on the ratio $l/N$. In the Mott case, a brute-force collapse of the correlations in the top panel of Figure~\ref{FIG_S9:  ML extrapolation correlations and data collapse} c) shows that data collapse is not possible (bottom panel of Figure~\ref{FIG_S9:  ML extrapolation correlations and data collapse} c)). Hence, implying that the predicted correlations in the Mott phase are not purely power-law. 
    
\begin{figure*}[t]
        \includegraphics[width=.8\textwidth]{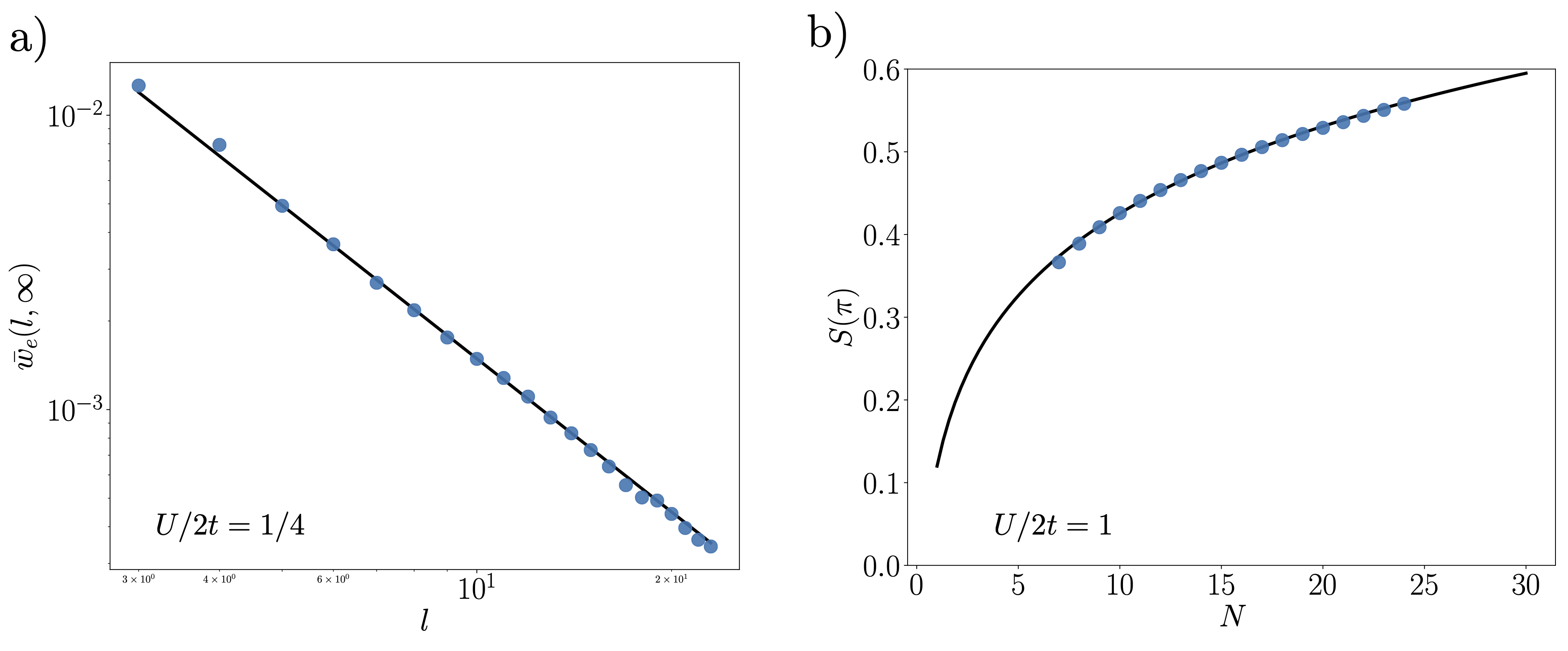}
        \caption{\label{FIG_S10:  ML extrapolated correlations+ log correction} Extrapolated correlation functions in the thermodynamic limit. LL parameter and logarithmic corrections in the critical point from ML correlations. \textbf{a)} Extrapolated correlation functions in the thermodynamic limit in the metallic phase. Blue dots represent the extrapolated values from Figure~\ref{FIG_S9:  ML extrapolation correlations and data collapse} a) and the black line to the fit to $a/l^{2K}$ where a and $K$ are the fitting parameters. \textbf{b)} Structure factor at $q = \pi$ in the critical point from extrapolated correlation functions in the thermodynamic limit from Figure~\ref{FIG_S9:  ML extrapolation correlations and data collapse} b) (blue dots). Black line is the fit to $S(\pi) = a + b\frac{2}{3}\log^{3/2}(cN)$, where $a$ and $b$ are the fitting parameters.} 
    \end{figure*}

    Finally, with the extrapolated values of the correlation functions in the thermodynamic limit we obtain the LL parameter and logarithmic corrections in the critical point. Figure~\ref{FIG_S10:  ML extrapolated correlations+ log correction} a) shows the extrapolated correlation functions in the thermodynamic limit as a function of $l$. They are consistent with a pure power-law decay. We can extract the LL parameter from the fit of such correlation functions and obtain $2K_\textrm{fit} = 1.73$, versus the exact value at this interaction strength $2K_\textrm{exact} = 1.7228$. This is a very accurate prediction of the LL parameter. The small deviation comes from the small errors in the ML correlations.
    
    Regarding the logarithmic corrections in the critical point, we look at the structure factor at $q = \pi$ as a function of chain length:
    \begin{equation}\label{Eq: Structure factor}
        S(\pi, N) = \sum_{l=1}^{N}\bar{w}_e(l,\infty)
    \end{equation}
    If $\bar{w}(l,\infty) \sim \sqrt{\log(cl)}/l$, then $S(\pi, N) = a +b \frac{2}{3}\log^{3/2}(cN)$, where it is known from the DMRG analysis that $c = 21.5$ and $a$ and $b$ are constants. Figure~\ref{FIG_S10:  ML extrapolated correlations+ log correction} b) shows the structure factor as a function of $N$ in the critical point and its fit to the expected value of the integral. There is a very good agreement between the fit and the extrapolated data, proving that we can capture the logarithmic corrections in the critical point.
    
    In conclusion, the ML-predicted correlation pairs accurately predict the decay scaling in the different phases of the system. Data collapse is possible in the metallic phase and critical point as the correlations are purely power-law and power-law with a small logarithmic correction respectively. The data collapse was not possible in the Mott phase, proving that correlation functions do not decay following a power-law. The correct LL  parameter was predicted in the metallic phase and the logarithmic corrections could be captured from the extrapolated correlation functions in the thermodynamic limit. 


%